\documentclass[twocolumn,floatfix,nofootinbib,here,showpacs,preprintnumbers,amsmath,amssymb]{revtex4}

\usepackage{graphicx}
\usepackage{dcolumn}
\usepackage{bm}

\begin{document}

\preprint{APS/123-QED}

\title{ Experimental study of  pp$\eta$ dynamics in the $pp\to pp\eta$ 
        reaction\protect\footnote{
            We are very pleased to dedicate the present paper
            to Prof.~Dr.~Dr.h.c.~Adam~Strza{\l}kowski on the 
            occasion of his 80$^{th}$ birthday
        } 
      }

\author{P.~Moskal$^{1,2}$,
H.-H.~Adam$^{3}$,
A.~Budzanowski$^{4}$,
R.~Czy$\dot{\mbox{z}}$ykiewicz$^{2}$,
D.~Grzonka$^{1}$,
M.~Janusz$^{2}$,
L.~Jarczyk$^{2}$,
B.~Kamys$^{2}$,
A.~Khoukaz$^{3}$,
K.~Kilian$^{1}$,
P.~Kowina$^{1,5}$,
K.~Nakayama$^{6}$,
W.~Oelert$^{1}$,
C.~Piskor-Ignatowicz$^{2}$,
J.~Przerwa$^{2}$,
T.~Ro$\dot{\mbox{z}}$ek$^{1,5}$,
R.~Santo$^{3}$,
G.~Schepers$^{1}$,
T.~Sefzick$^{1}$,
M.~Siemaszko$^{5}$,
J.~Smyrski$^{2}$,
S.~Steltenkamp$^{3}$,
A.~T{\"a}schner$^{3}$,
P.~Winter$^{1}$,
M.~Wolke$^{1}$,
P.~W{\"u}stner$^{7}$,
W.~Zipper$^{5}$
}

\affiliation{$^1$ Institut f{\"u}r Kernphysik, Forschungszentrum J\"{u}lich, D-52425 J\"ulich, Germany}
\affiliation{$^2$ Institute of Physics, Jagellonian University, PL-30-059 Cracow, Poland}
\affiliation{$^3$ Institut f{\"u}r Kernphysik, Westf{\"a}lische Wilhelms--Universit{\"a}t, D-48149 M{\"u}nster, Germany}
\affiliation{$^4$ Institute of Nuclear Physics, PL-31-342 Cracow, Poland}
\affiliation{$^5$ Institute of Physics, University of Silesia, PL-40-007 Katowice, Poland}
\affiliation{$^6$ Department of Physics and Astronomy, 
                  University of Georgia, Athens, GA 30602, USA}
\affiliation{$^7$ Zentrallabor f{\"u}r Elektronik, 
                Forschungszentrum J\"{u}lich, D-52425 J\"ulich, Germany}
\date{\today}

\begin{abstract}
 A high statistics  measurement of the $pp\rightarrow pp\eta$ reaction
 at an excess energy of Q~=~15.5~MeV  has been performed at
 the internal beam facility COSY-11.
 The stochastically cooled proton beam and the used detection system
 allowed to determine the momenta of the outgoing protons with a precision
 of 4~MeV/c~($\sigma$) in the center-of-mass frame.
 The determination of the four-momentum vectors of both outgoing protons
 allowed to 
 derive the complete kinematical
 information of the $\eta$pp-system.
 An unexpectedly large enhancement 
 of the occupation density in the kinematical regions of low
 proton-$\eta$ relative momenta is observed.
 A description taking the proton-proton and  the $\eta$-proton interaction into account 
 and assuming an on-shell
 incoherent pairwise interaction among the produced particles fails to explain this strong effect.
 Its understanding 
 will require
 a rigorous three-body approach to the $pp\eta$ system and the precise determination
 of contributions from higher partial waves.
 We also present an invariant mass spectrum of the proton-proton system
 determined at Q~=~4.5~MeV.
 Interestingly, the enhancement at large relative
 momenta between protons is visible also at such a small excess energy.

 In contrast to all other determined angular distributions,
 the orientation of the emission plane with respect
 to the beam direction is extracted to be anisotropic.
\end{abstract}

\maketitle

\vspace{0.1cm}

\section{Manifestation of the $\eta$-p-p interaction}
  Due to the short live time of the flavour-neutral 
  mesons~(eg. $\pi^0$, $\eta$, $\eta^{\prime}$), the study of their 
  interaction with nucleons or with other mesons is at present not feasible
  in direct scattering experiments. One of the methods permitting such 
  investigations is the production of a meson in the nucleon-nucleon
   interaction close to the kinematical threshold or in kinematics regions
  where the outgoing particles possess small relative velocities.
  A mutual  interaction among the outgoing particles
  manifests itself in the distributions
  of differential cross sections as well as in 
  the magnitude and energy dependence
  of the total reaction rate.  
 
  In the last decade  major
  experimental~\cite{etap_data,hibou,eta_data,bergdoldeta,caleneta,
                     smyrskieta,pi0_data,meyerpolar}
  and theoretical~\cite{eta_theo,vadimm,nakayamaa,etap_theo,kaiser} efforts
  were concentrated on the study of the creation of
  $\pi^{0}$, $\eta$ and  $\eta^{\prime}$ mesons
  via the hadronic interactions~\cite{review,wilkinjohansson,machnerhaidenbauer}.
   Measurements 
  have been performed
  in the vicinity of the kinematical threshold where only 
  a few partial waves in both
  initial and final state
  are expected to contribute to the production process.
    This simplifies significantly the
  interpretation of the data, yet still appears to be challenging
  due to the three particle final state system with a complex hadronic potential.
  \begin{figure}[htb]
    \includegraphics[width=0.4\textwidth]{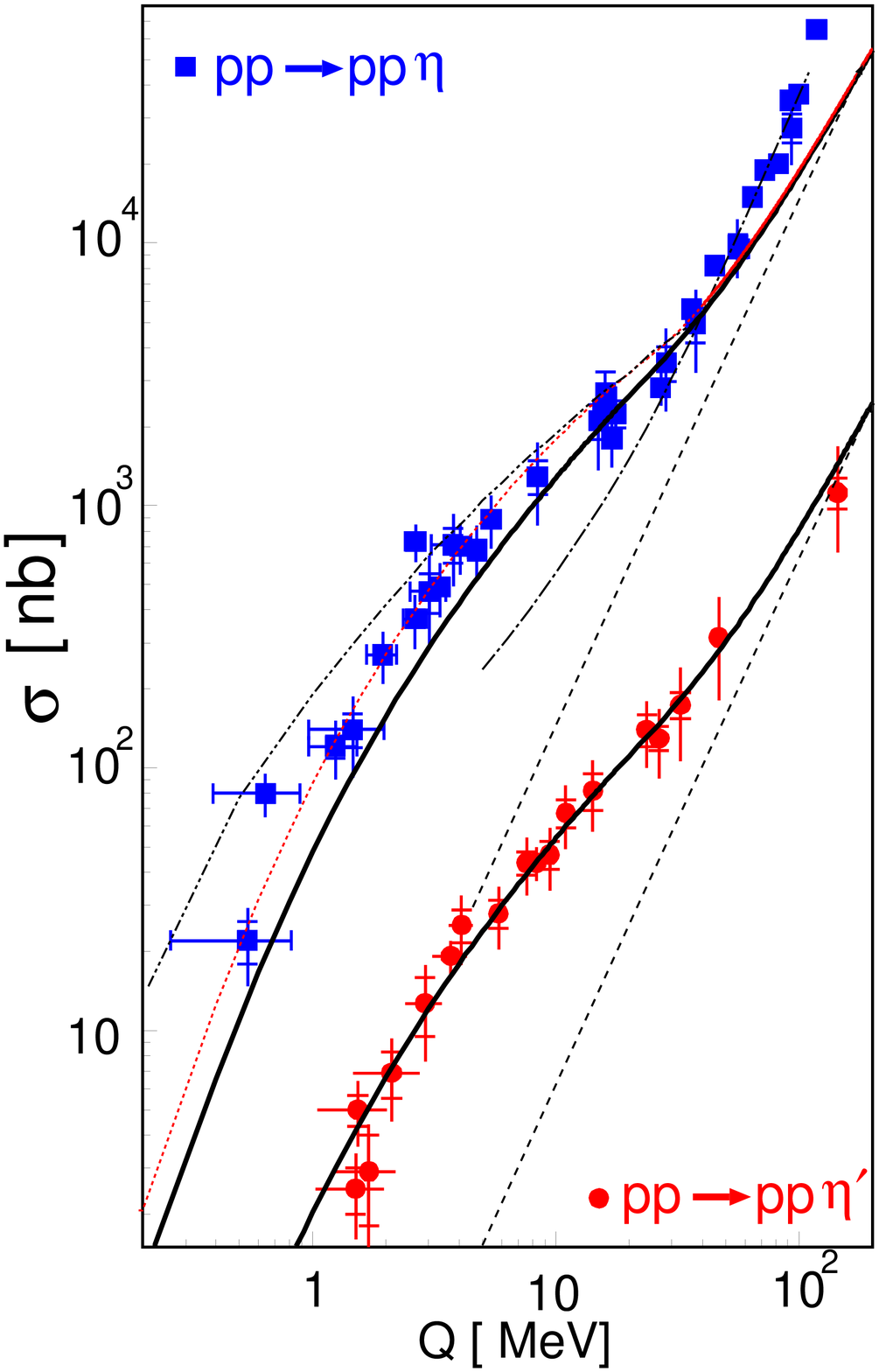}
    \vspace{-0.3cm}
    \caption{\label{cross_eta_etap}
      Total cross section for the reactions
      $pp \rightarrow pp \eta^{\prime}$~(circles) and 
      $pp \rightarrow pp \eta$~(squares)
      as a function of the centre--of--mass
      excess energy Q. Data are from
      refs.~\cite{etap_data,hibou,eta_data,bergdoldeta,caleneta,smyrskieta}.
      The dashed lines
      indicate a phase--space integral normalized arbitrarily. \\
      The solid lines
      show the phase--space distribution with inclusion of the 
      $^1S_0$ proton--proton strong
      and Coulomb interactions. 
    In case of the $pp\to pp\eta$ reaction the solid line 
    was fitted to the data in the excess energy range
    between 15 and $40\,\mbox{MeV}$. Additional
    inclusion of the proton--$\eta$ interaction is indicated by the dotted line.
    The scattering length of $a_{p\eta} = 0.7\,\mbox{fm} + i\,0.4\,\mbox{fm}$ and the
    effective range parameter $r_{p\eta} = -1.50\,\mbox{fm} -
    i\,0.24\,\mbox{fm}$~\cite{greenR2167} have been arbitrarily chosen.
    The dashed-dotted line represents the energy dependence taking into account the 
    contribution from the   
    $^3P_{0}\to ^1\!\!\!S_{0}s$, $^1S_{0}\to ^3\!\!P_{0}s$ and 
    $^1D_2\to ^3\!\!P_2 s$ transitions~\cite{nakayamaa}.
    Preliminary results for the $^3P_{0}\to ^1\!\!S_{0}s$
    transition with the full treatment of the three-body effects
    are shown as a dashed-double-dotted line~\cite{Fixprivate}.
    The absolute scale of dashed-double-dotted line 
    was arbitrary fitted to demonstrate 
    the energy dependence only.
    }
    \vspace{-0.6cm}
  \end{figure}
  The determined energy dependences of the total cross section
  for $\eta^{\prime}$~\cite{etap_data,hibou} and 
  $\eta$~\cite{hibou,eta_data,bergdoldeta,caleneta,smyrskieta}
  mesons in  proton-proton collisions
  are presented in figure~\ref{cross_eta_etap}.
  Comparing the data to the arbitrarily normalized phase-space integral (dashed lines)
  reveals that the proton-proton FSI enhances the total cross section by more than an order
  of magnitude for low excess energies.
  One recognizes also that
  in the case of the $\eta^{\prime}$
  the data are described very well (solid line)
    assuming that the on-shell proton-proton amplitude
    exclusively determines the phase-space population.
  This indicates that the  proton-$\eta^{\prime}$ interaction is too small to manifest itself
  in the excitation function within the presently achievable accuracy.
  In case of the $\eta$ meson the increase of the
  total cross section for very low and very high energies is much larger than expected
  from the  $^1S_0$ final state interaction between protons~(solid line),
  though  for both the $pp\to pp\eta$ and $pp\to pp\eta^{\prime}$ reactions
  the dominance
  of the $^3P_{0}\to ^1\!\!S_{0}s$ 
  transition\footnote{
   The transition between angular momentum combinations of the initial and final
   states are described according to the conventional notation~\cite{meyerpolar}
   in the following way:
   \begin{equation}
        ^{2S^i+1}L_{J}^i \rightarrow ^{2S+1}\!\!L_{J},l
   \end{equation}
   where, superscript "i" indicates the initial state quantities.
   S denotes the total spin of the nuleons,  and
   J stands for the overall angular momentum of the system.
   $L$ and $l$ denote the relative angular momentum of the nucleon--nucleon
   pair and of the meson relative to the $NN$ system, respectively.
   The values of orbital angular momenta
   are commonly expressed using the spectroscopic notation
   (L~=~S,P,D,..., and l~=~s,p,d,...).
  }
  is expected up to an excess energy
  of about 40~MeV and 100~MeV, 
  respectively~\cite{review}.
  The excess at higher energies
  can be assigned to the significant onset of higher partial waves, and
  the influence of the attractive
  interaction between the $\eta$ meson and the proton 
  could be a plausible explanation for the enhancement at threshold.
  A similar effect close to threshold is also observed in the photoproduction of $\eta$
  via  the $\gamma d\to pn\eta$ reaction~\cite{eta_photo} indicating
  to some extent that the phenomenon is independent of the production
  process but rather related to the interaction among the $\eta$ meson and
  nucleons in the S$_{11}$(1535) resonance region.
  Indeed, a simple phenomenological treatment~\cite{review,kaiser,szubert}
   --~based on factorization of the 
  transition amplitude into the constant primary production and the on-shell
  incoherent pairwise interaction among outgoing particles~-- describes very well
  the enhancement close to the threshold~(dotted line).
  However, this approach fails for the description of the invariant 
  mass distribution of the proton-proton and proton-$\eta$ subsystems
  determined recently at Q~=~15~MeV by the COSY-TOF collaboration~\cite{TOFeta}. 
   The structure of this invariant mass distributions, 
   which we confirm in this paper utilizing a fully different experimental method,  
  may indicate  a non-negligible contribution from the P-waves in the 
  outgoing proton-proton subsystem~\cite{nakayamaa}.
  These can be produced for instance via  $^1S_{0}\to ^3\!\!P_{0}s$ or $^1D_2\to ^3\!\!P_2 s$
  transitions. This hypothesis encounters, however, difficulties in describing the 
  excess energy dependence of the total cross section.
  The amount of the P-wave admixture  derived from the proton-proton invariant mass
  distribution  leads to a good description of the excitation function 
  at higher excess energies
  while at the same time  it spoils significanly 
  the agreement with  the data at low values of Q, as depicted by the 
  dashed-dotted line in figure~\ref{cross_eta_etap}. 
  However, these difficulties in reproducing the observed energy dependence
  might be due to the particular model used in reference~\cite{nakayamaa},
  and thus higher partial wave contributions cannot a priori be excluded.
  In contrast to the P-wave contribution
  the three-body treatment~\cite{Fixprivate} of the $pp\eta$
  system~(dashed-double-dotted line) 
  leads to an even larger enhancement of the cross section near threshold 
  than that based on 
  the Ansatz of the factorization of the proton-proton and proton-$\eta$ interactions.
  It must be kept in mind however that a too strong FSI effect predicted by the three-body
  model must be partially assigned to the neglect of the Coulomb repulsion in this
  preliminary calculations~\cite{Fixprivate}.
  These illustrate that the 
  simple phenomenological approach shown by the dotted line 
  could fortuitously lead to the proper result, due to a mutual cancelation
  of the effects caused by the approximations assumed in calculations and
  the neglect  of higher partial waves. 
  This issue will be discussed further in section~\ref{interpretation} after the
  presentation of the new COSY-11 data.
  The above considerations show unambiguously that for the complete understanding
  of the low energy $pp\eta$ dynamics, in  addition to the already established
  excitation function of the total cross section, a determination of the
  differential observables is also necessary. 
\begin{figure*}[t]
    \parbox{1.0\textwidth}{ 
    \includegraphics[width=0.85\textwidth]{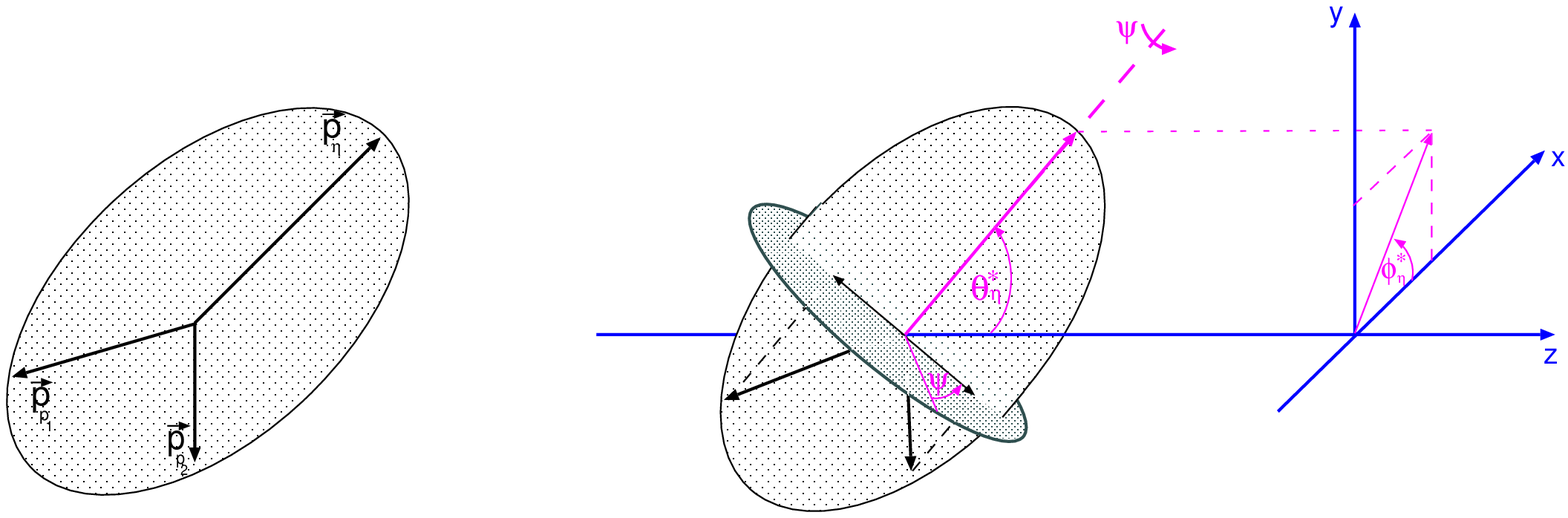}}
    \parbox{1.0\textwidth}{ \vspace{-0.4cm}
    \includegraphics[width=0.85\textwidth]{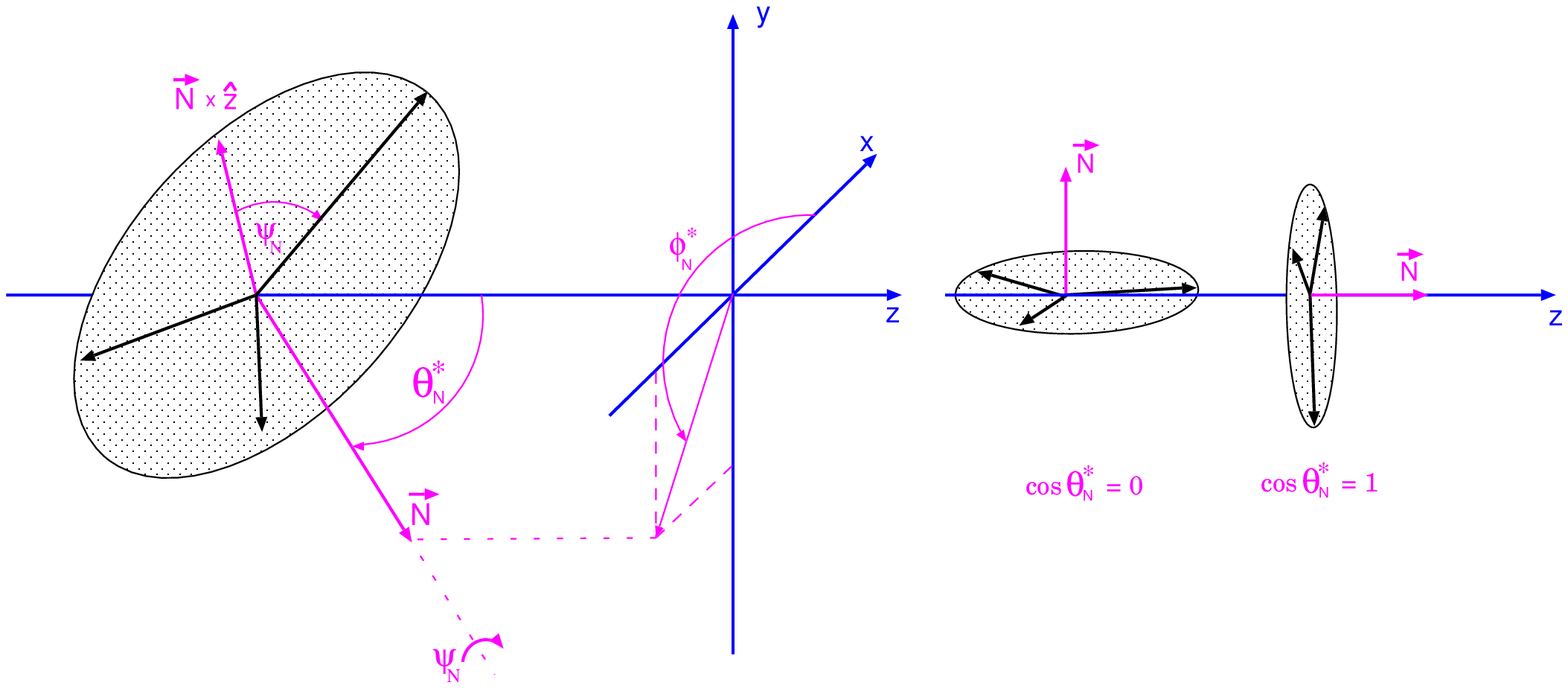}}
    \vspace{-0.3cm}
   \caption{ \label{figvariables}
     Definition of the centre-of-mass kinematical variables used in this article
     for the description of the pp$\eta$ system.
     In the centre-of-mass system, the momenta of ejectiles lie in an emission plane.
     Within this plane the relative movement of the particles
     is fixed by the square of the invariant masses $s_{pp}$ and $s_{p\eta}$.
     As  remaining three variables needed to define the system uniquely
     we use 
     either $\phi^{*}_{\eta}$, $\theta^{*}_{\eta}$, and $\psi$
     shown in the upper panel or $\phi^{*}_{N}$, $\theta^{*}_{N}$, and $\psi_{N}$
     defined in the lower panel.
     $\vec{N}$ is a vector normal to the emission plane, which can be calculated 
     as the vector product of the centre-of-mass momentum vectors of the outgoing protons.
     As an example two extreme orientations
     of the emission plane are shown in the right-lower panel.
     For  further descriptions see the text.
   }
\end{figure*}
  These will help to disentangle effects
  caused by the proton-$\eta$  interaction and the  contributions
  from higher partial waves. 
  In this article we present distributions determined experimentally  
  for two sets of orthogonal variables
  fully describing the pp$\eta$  system, which was produced at an excess energy of Q~=~15.5~MeV
  via the $pp\to pp\eta$ reaction using the COSY-11~\cite{brauksiepe,nim} 
  facility at COSY~\cite{prasuhn}.
\section{Choice of observables}
\label{choiceofobservables}
For the full description of the three particle system five independent variables are required.
In the center--of--mass frame,
due to the momentum conservation, 
the momentum vectors of the particles are lying in one  plane often referred 
to as the emission,
reaction or decay plane. In this plane~(depicted in figure~\ref{figvariables})
a relative movement of the particles can be described by two variables only. 
The square of the invariant masses of the di-proton and proton-$\eta$ 
system denoted as $s_{pp}$ and $s_{p\eta}$, respectively,
constitute a natural choice for the study 
of the interaction within  the pp$\eta$ system.
This is because in the case of non-interacting objects
the surface spanned by these variables is homogenously populated. 
The interaction among 
the particles modifies that occupation density
and in consequence facilitates an easy 
qualitative  interpretation of the 
experimental results.

The remaining three variables must define an absolute 
orientation of the emission plane in the distinguished coordinate system. 
This may be realized for example by defining the orientation 
for the momentum of the arbitrarily chosen particle
in the center--of--mass frame
and the angle which describes the rotation around the direction fixed by that particle.
In one of our choices made following reference~\cite{balestra,jim}
the corresponding variables are the polar and the azimuthal
angle of the $\eta$ momentum vector, depicted in figure~\ref{figvariables}
as $\phi^*_{\eta}$ and $\theta^{*}_{\eta}$, respectively,  and the angle $\psi$ describing 
the rotation around the direction established by the momentum of the $\eta$ meson. 
Figure~\ref{figvariables} demonstrates that such rotation neither affects the $\eta$ meson
momentum nor  changes the configuration
of the momenta in the emission plane.
In the case of experiments with  unpolarized beams and targets the only favoured direction 
is that of the beam line. Therefore, as a  zero value of the $\psi$ angle we have chosen the 
projection of the beam direction 
on the plane perpendicular to the momentum vector of the $\eta$
meson. Note that we identify the z-axis with the beam direction. 
The $\phi^*_{\eta}$, $\theta^{*}_{\eta}$ and $\psi$ variables 
can also be interpreted as Euler angles allowing  for the 
rotation of the emission plane into a xz-plane.

As a second possibility we will describe an orientation of the emission plane by the 
azimuthal- and polar angle of the vector normal to that plane~\cite{Eduardacta}.
These angles are
shown in figure~\ref{figvariables} as $\phi^{*}_{N}$ and $\theta^{*}_{N}$, 
respectively.  Further  the absolute orientation 
of the particles momenta in the emission plane will be described by $\psi_{N}$,
the angle between $\eta$ meson and the vector product of the beam momentum
and the vector $\vec{N}$.

Obviously, the interaction between the particles does not depend on the orientation 
of the emission plane, and therefore, it will fully manifest itself in the 
occupation density of the Dalitz plot which in our case will be represented in terms
of the square of the invariant masses of the two particle subsystems.
The distribution of the orientation of the emission plane will reflect, however,
the correlation between the intial and final channel and hence its determination
should be helpful for the investigation of the production mechanism.

\section{Experiment and data analysis}

 Using the COSY-11 detection system~\cite{brauksiepe,nim},
 utilizing  a stochastically cooled proton beam
 of the cooler synchrotron COSY~\cite{prasuhn}
 and a hydrogen cluster target~\cite{dombrowski},
 we have performed a high
 statistics measurement of the $pp \rightarrow pp\eta$ reaction
 at a nominal beam momentum of 2.027~GeV/c.
  \begin{figure}[t]
    \parbox{0.5\textwidth}{\vspace{-0.4cm}
    \includegraphics[width=0.48\textwidth]{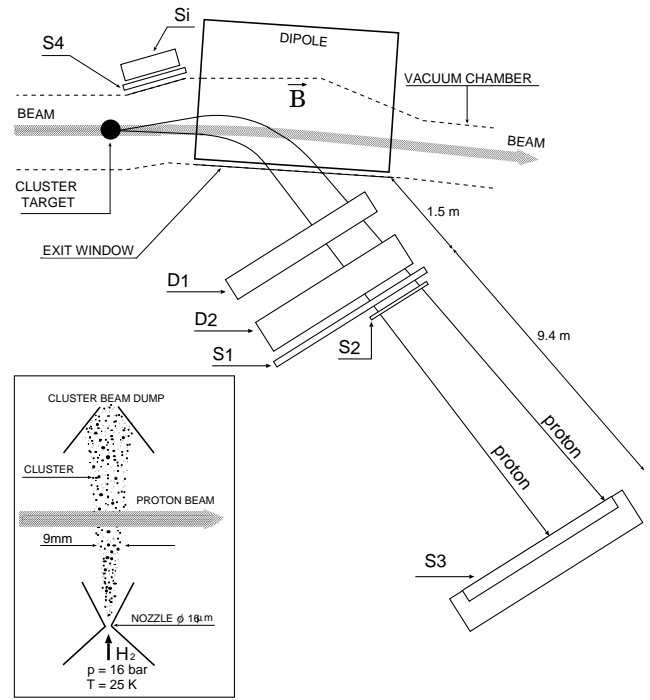}}
    \vspace{-0.3cm}
    \caption{\label{detector}
       Schematic view of the COSY-11 detection setup~\cite{brauksiepe}.
       The cluster target~\cite{dombrowski} is located in front 
       of the accelerator dipole magnet. 
       Positively charged particles which leave the
       scattering chamber through the thin exit foil are detected
       in two drift chamber stacks D1, D2 and in the scintillator hodoscopes
       S1, S2 and S3.
       Scintillation detector S4 and the 
      position sensitive silicon pad detector Si
      are used in coincidence with the S1 counter for the registration  of the elastically
      scattered protons. Elastic scattering is  used for an absolute normalisation 
      of the cross sections of the 
      investigated reactions and for monitoring both the geometrical spread of the proton 
      beam and the position at which  beam crosses the target~\cite{nim}. \ \ \ 
       In the left-lower corner  a schematic view of the interaction region
       is depicted. 
    }
    \vspace{-0.2cm}
  \end{figure}  
 The experiment was based on the four-momentum registration
 of both outgoing protons, whereas the $\eta$ meson was identified
 via the missing mass technique. The positively charged particles 
 have been identified  
 combining the time of flight between the S1 and S3 
 scintillation detectors and the momentum reconstructed 
 by tracking trajectories registered 
 by means of the drift chambers back to the target. 
 The detection setup is sketched
 in figure~\ref{detector}.
 
\begin{figure}[h]
  \parbox{0.44\textwidth}{\vspace{-1.6cm}
     \includegraphics[width=0.44\textwidth]{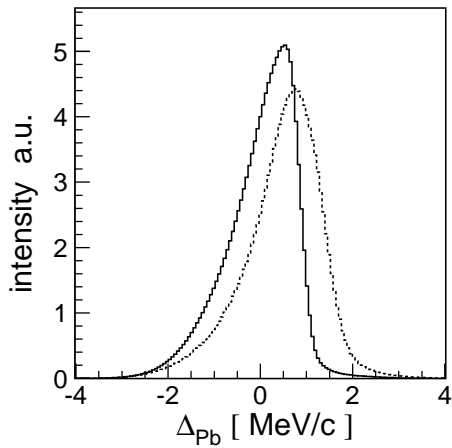}}
  \vspace{-0.7cm}
      \caption{ The dashed curve denotes the 
           proton beam momentum distribution 
           integrated over the whole measurement period.
           On the horizontal scale, the value of zero is set at a nominal beam momentum
           equal to 2.027~GeV/c.
           The solid line shows the
           beam momentum distribution after the correction for
           the mean value which was
           determined in  10 second intervals. 
           \label{delta_Pb}
        }
\end{figure}

 The accuracy of the missing mass reconstruction
 depends on two parameters: the spread of the beam momentum as well as the 
 precision of the measured momenta of the outgoing protons. The latter is 
 predominantly due to the geometrical spread of the beam.  
 Since for the reconstruction of the momenta we assume that the reaction
 takes place in the middle of the target we cannot correct 
 on an event-by-event basis for the momentary spread of the beam.
 However, we can rectify the smearing due to the shifts of the centre of the beam relative
 to the target  as well as the average changes of the absolute beam momentum 
 during the experiment.
Therefore, after the selection of events with two registered protons,
as a first step of the more refined analysis
the data were corrected for the mean beam momentum
changes (see fig.~\ref{delta_Pb}) determined from the
measured Schottky frequency spectra and the known beam optics.
In a next step, from the distributions of the elastically scattered protons,
the Schottky frequency spectrum, and the missing mass distribution
of the $pp \to pp X$ reaction,
we have estimated that 
the spread of the beam momentum,
and the spread of the reaction points in horizontal and
vertical direction
amount to
$\sigma(p_{beam})~=~0.63~\pm~0.03$~MeV/c,
$\sigma(x)~=~0.22~\pm~0.02$~cm,
and $\sigma(y)~=~0.38~\pm~0.04$~cm, respectively.
Details of this procedure can be found in references~\cite{nim,phd}.
\begin{figure}
  \parbox{0.5\textwidth}{\vspace{-1.5cm}
    \includegraphics[width=0.4\textwidth]{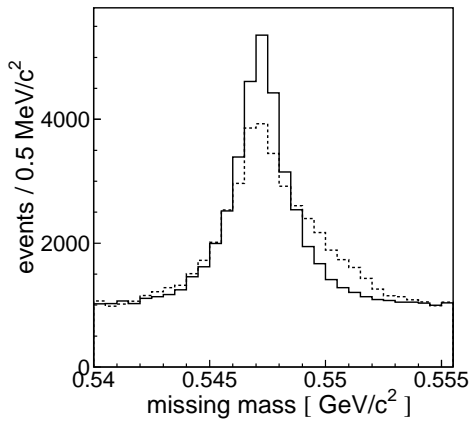}}
  \vspace{-0.8cm}
  \caption{ Missing mass distribution for the $pp\rightarrow ppX$
           reaction determined by means of the COSY-11 detection system
           at a beam momentum of 2.027~GeV.
          The solid histogram presents the data corrected
          for effects of the time dependent relative
          shifts between the beam and
          the target using the method described in reference~\cite{nim}.
          The dashed histogram shows the result before the correction.
           \label{miss_all}
        }
\end{figure}
Further on, a comparison of the experimentally determined
momentum spectra of the elastically scattered protons
with the distributions simulated with different beam and target
conditions allows us to established the position at which the
centre of the 
beam crosses the target with an accuracy of 0.25~mm~\cite{nim}.
Accounting for the movement of the beam
relative to the target 
we improved the missing mass resolution
as demonstrated in figure~\ref{miss_all}.
\begin{figure}[t]
  \parbox{0.5\textwidth}{\vspace{-1.1cm}
    \includegraphics[width=0.43\textwidth]{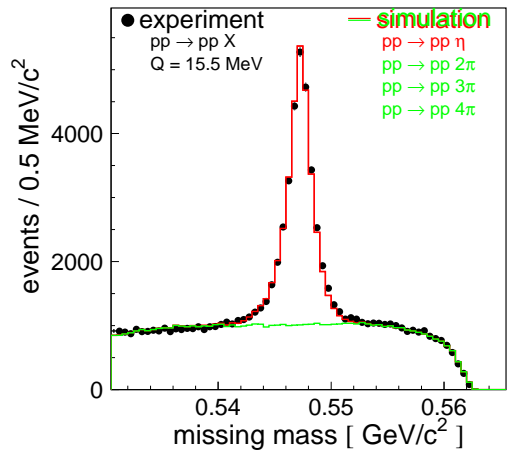}}
  \vspace{-0.9cm}
  \caption{ \label{missall} 
         Missing mass spectrum for the $pp\to ppX$ reaction
         determined in the experiment at a beam momentum of 2.0259~GeV/c.
         The mass resolution amounts to 1~MeV/c$^2(\sigma)$.
         The superimposed histograms present the simulation
         for $1.5\cdot 10^8$ events of the $pp\to pp\eta$ reaction,
         and $10^{10}$ events for the reactions $pp\to pp 2\pi$,
         $pp\to pp 3\pi$ and $pp\to pp 4\pi$.
         The simulated histograms were fitted to the data varying only the magnitude.
         The fit resulted in $24009\pm 210$ events with the production
         of the $\eta$ meson.
 } 
\end{figure}
After this betterment, the peak originating from the $pp\to pp\eta$ reaction became 
 more symmetric and the signal to background ratio increased significantly.
The background was simulated taking into account $pp\to pp X$ reactions with
$X~=~2\pi,~3\pi$ and~$4\pi$.
Since we consider here only
the very edge of the phase space distribution where the protons are
produced predominantly in the S-wave the shape of the background can
be reproduced assuming that the homogenous phase space distribution
is modified only by the interaction between protons. Indeed, as can be observed
in figure~\ref{missall} the simulation describes the data very well.
The calculated spectrum is hardly distinguishable from the experimental points.

The position of the peak on the missing mass spectrum
and the known mass of the $\eta$ meson~\cite{PDG}
enabled to determine the actual absolute beam
momentum to be $p_{beam}~=~2.0259$~GeV/c~$\pm$~0.0013~GeV/c,
which agrees within error limits with the 
nominal value of $p_{beam}^{nominal}~=~2.027$~GeV/c.
The real beam momentum corresponds to the excess energy of the pp$\eta$ system 
equal to Q~=~15.5~$\pm$~0.4~MeV.

\subsection{Covariance matrix and kinematical fitting}
\begin{figure}[t]
  \parbox{0.5\textwidth}{\vspace{-0.3cm}
    \includegraphics[width=0.35\textwidth]{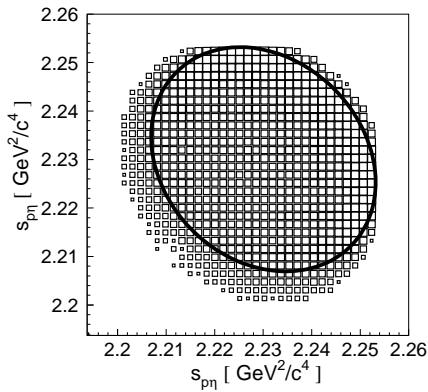}}
  \vspace{-0.35cm}
  \caption{Dalitz plot distribution of the $pp\to pp\eta$ reaction
         simulated at Q~=~15.5~MeV.
         The number of entries is shown
          in a logarithmic scale.
         The solid line gives the kinematically allowed area.
          The  result was obtained
          taking into account the experimental conditions
          as described in the text.
         \label{dalitzsym}
  }
\end{figure}

 As already mentioned in the previous section
 at the COSY-11 facility the identification of the $pp \to pp \eta$
reaction is based on the measurement of the momentum vectors of
the outgoing protons and the utilisation of the missing mass technique.
Inaccuracy of the momentum determination manifests itself in the
population of  kinematically forbidden regions of the phase space,
preventing a precise comparison of the theoretically derived
and experimentally determined differential cross sections. Figure~\ref{dalitzsym}
visualizes this effect and
clearly demonstrates that the data
scatter significantly
outside the kinematically allowed region~(solid line),
in spite of the fact that the precision of the fractional momentum determination
in the laboratory system ($\sigma(p_{lab})/p_{lab}\approx 7\cdot 10^{-3}$) is quite high.
Therefore, when seeking for  small effects
like for example the influence of the proton-$\eta$ interaction
on the population density of the phase-space,
one needs either to fold theoretical calculations with the experimental resolution,
or to perform  the kinematical fitting of the data.  Both procedures require the knowledge
of the covariance matrix, and thus its determination constitutes a necessary step
in the differential analysis and interpretation of the data.

In order to derive the covariance matrix we need to recognize and quantify
all possible sources of errors
in the reconstruction of the two proton momenta $\vec{p}_1$ and $\vec{p}_2$.
The four dominant effects are:
i)   finite distributions of the beam momentum and of the reaction points,   
ii)  multiple scattering in the dipole chamber exit foil, air, and detectors,
iii) finite resolution of the position determination of the drift chambers, and 
iv) a possible inadequate assignment of hits to the particle tracks in drift chambers in the case
    of very close tracks.
Some of these, like the multiple scattering, depend on the outgoing protons' momenta,
others, like the beam momentum distribution, depend on the specific
run conditions and therefore must be determined for each run separately.
 
In order to estimate the variances and covariances for all possible
combinations of the momentum components of two registered protons
we have generated $1.5\cdot 10^8$ $pp\to pp\eta$ events and simulated the response
of the COSY-11 detection setup taking into account the above listed
factors and the known resolutions of the detector components.
Next, we  analysed the signals by means of the same reconstruction procedure
as used in case of the experimental data.
Covariances between the $i^{th}$ and the $j^{th}$ components of the
event vector ${\displaystyle{ \left( P=[p_{1x},p_{1y},p_{1z},p_{2x},p_{2y},p_{2z}] \right)}}$
were established as the average of the product of the deviations between the reconstruced
and generated values.
The explicit formula for the sample of $N$ reconstructed events reads:
 
\vspace{-0.6cm}
 
\begin{equation}
   \nonumber
   cov(i,j)~=~\frac{1}{N} \sum_{k=1}^{N}(P_{i,gen}^k - P_{i,recon}^k)(P_{j,gen}^k - P_{j,recon}^k),
\end{equation}

\vspace{-0.2cm}

\hspace{-0.5cm} where $P_{i,gen}^k$ and $P_{i,recon}^k$ denote the generated and reconstructed
values for the $i^{th}$ component of the vector $P$ describing the $k^{th}$ event.
 
Because of the inherent symmetries of the covariance matrix 
$(cov(i,j)=cov(j,i))$
and the indistinguishability of the registered 
protons~\footnote{
  The symmetry of all observables under the exchange of the two protons 
  ($\vec{p}_1 \leftrightarrow \vec{p}_2$) implies that 
   cov(i,j)~=~cov(i$\pm$3,j$\pm$3), where the '+' has to be taken
   for i,j~=~1,2,3 and the '-' for i,j~=~4,5,6. Thus for example
   cov(2,4)~=~cov(5,1).}
there are only
12 independent values which determine the $6~{\mbox{x}}~6$
error matrix $V$ unambiguously.
 
Since inaccuracies of the momentum determination depend on the particle momentum
itself (eg.~multiple scattering) and on the relative momentum between protons
(eg. trajectories reconstruction from signals in drift chambers),
we have determined the covariance matrices as a function of the  absolute momentum
of both protons: $cov(i,j,|\vec{p_{1}}|,|\vec{p_{2}}|)$.
\begin{figure}[t]
  \parbox{0.4\textwidth}{\vspace{-1.5cm}
    \includegraphics[width=0.253\textwidth]{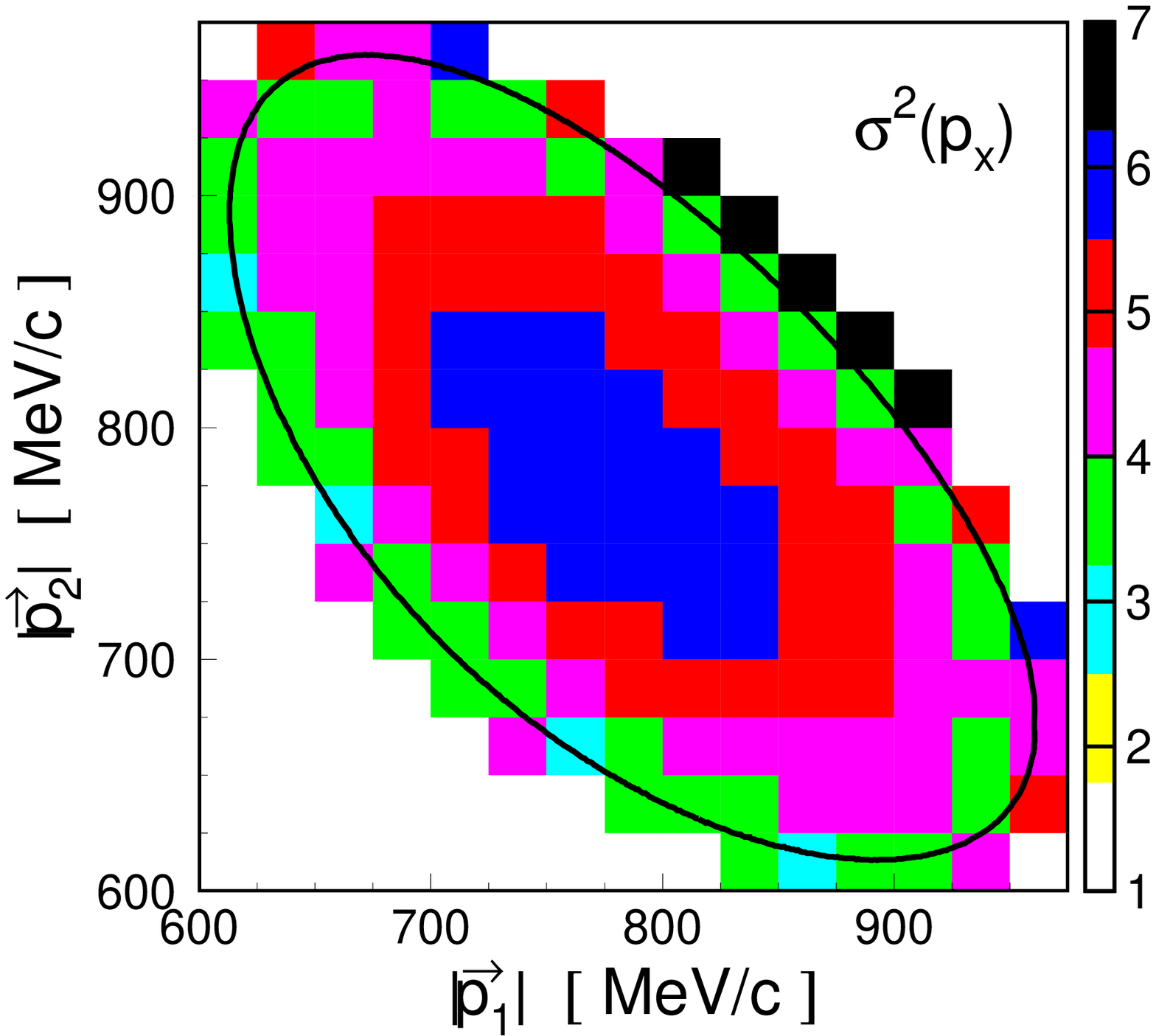}}
  \parbox{0.4\textwidth}{\vspace{-0.90cm}
    \includegraphics[width=0.253\textwidth]{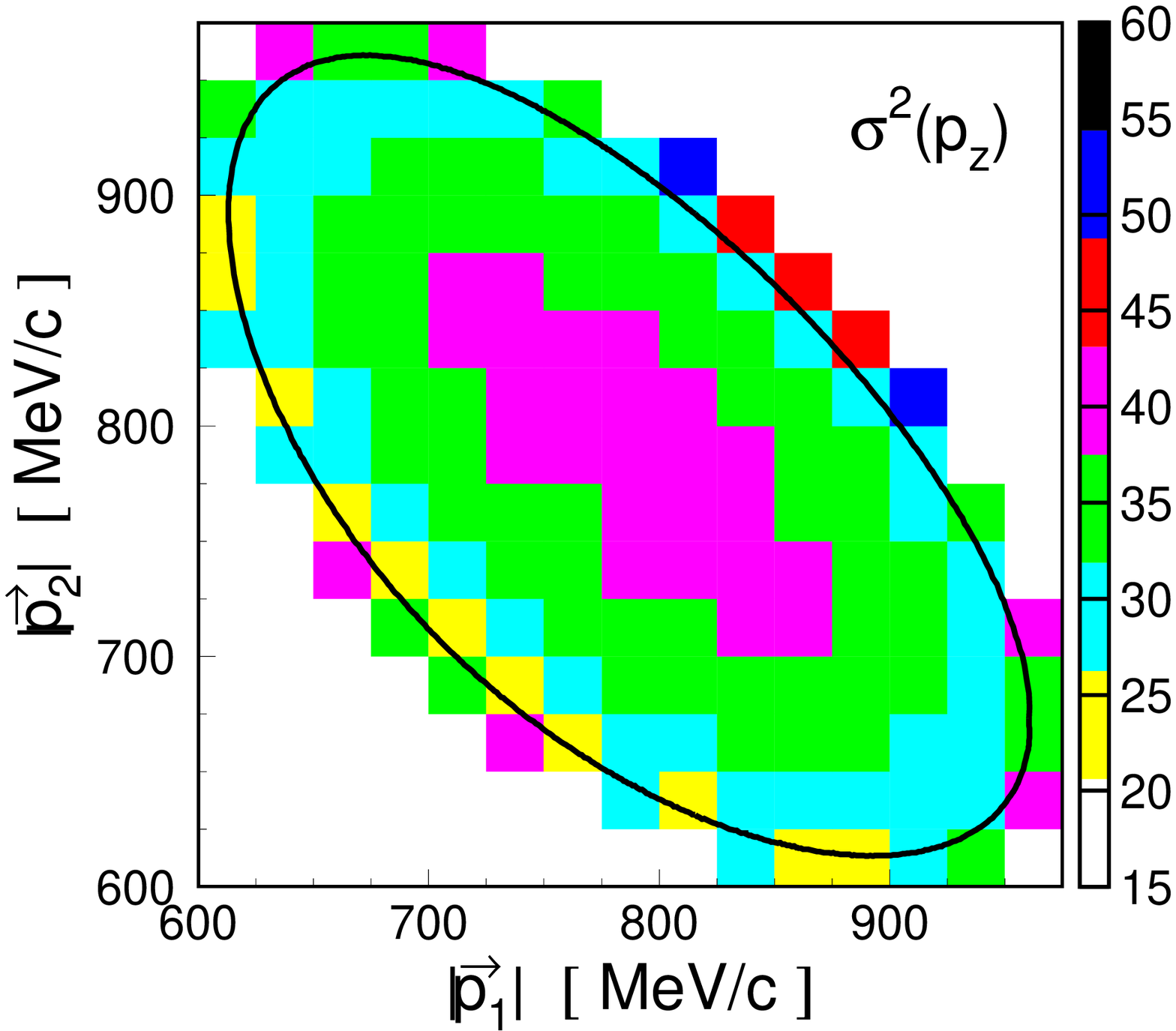}}
  \parbox{0.4\textwidth}{\vspace{-0.90cm}
    \includegraphics[width=0.253\textwidth]{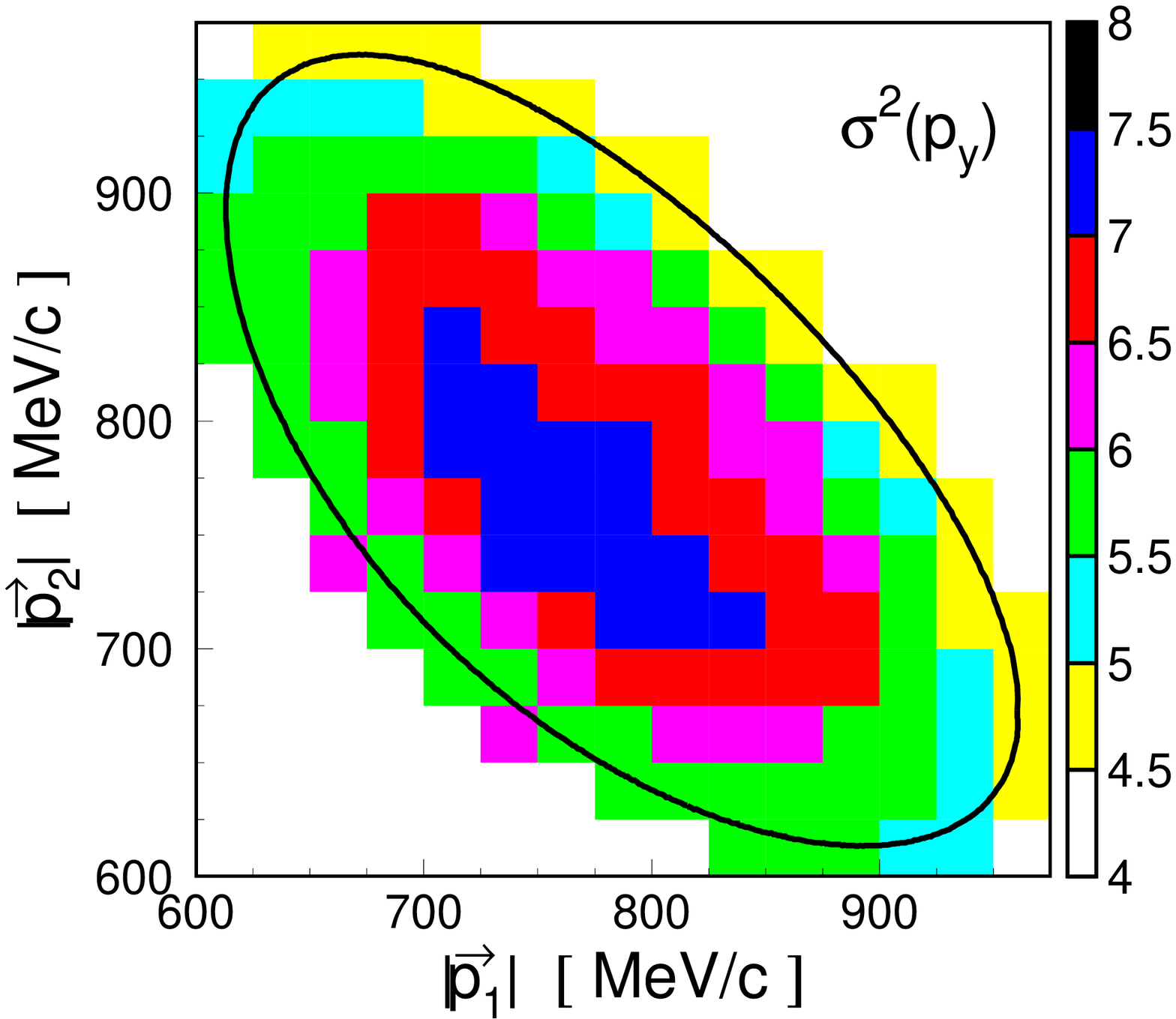}}
   \vspace{-0.2cm}
  \caption{ \label{covariance}
     Variances of the protons's momentum components 
     $\sigma^2(p_x)$, $\sigma^2(p_z)$, and $\sigma^2(p_y)$ shown
     as a function of the absolute values of the measured momenta.
  }
\end{figure}
 
As an example we present
the covariance matrix for the mean values of $|\vec{p_{1}}|$
and  $|\vec{p_{2}}|$ in units of MeV$^2$/c$^2$,
as established in the laboratory system with  the $z$-coordinate parallel
to the beam axis and $y$-coordinate corresponding to the vertical direction.
 
\vspace{-0.7cm}
\hspace{-1cm}\begin{equation}
 \mbox{}\hspace{-0.7cm}V~=~\begin{array}{c}
 \begin{array}{cccccc}
       p_{1x} \ \  &   p_{1y} \ \ &  p_{1z} \ \ & p_{2x} \ \ &  p_{2y} \ \ &  p_{2z}
 \end{array}
 \begin{array}{c}
  \mbox{\ \hspace{0.3cm}\ }
 \end{array}\\
 \left[
   \begin{array}{cccccc}
       5.6   &   0.0 &  -13.7 & 1.7 & 0.1  &  -3.0 \\
       -     &   7.1 &  0.1 &  - & -0.2  &  -0.2 \\
       -     &   -   &  37.0 & - & -  &  5.4 \\
       -   &   - &  - & - & -  &  - \\
       -   &   - &  - & - & -  &  - \\
       -   &   - &  - & - & -  &  -
   \end{array}
 \right]
 \begin{array}{c}
      p_{1x} \ \\\
      p_{1y}\\
      p_{1z}\\
      p_{2x}\\
      p_{2y}\\
      p_{2z}
 \end{array}
 \end{array}
\end{equation}
 
\begin{figure}[t]
  \parbox{0.44\textwidth}{\vspace{-1.1cm}
    \includegraphics[width=0.44\textwidth]{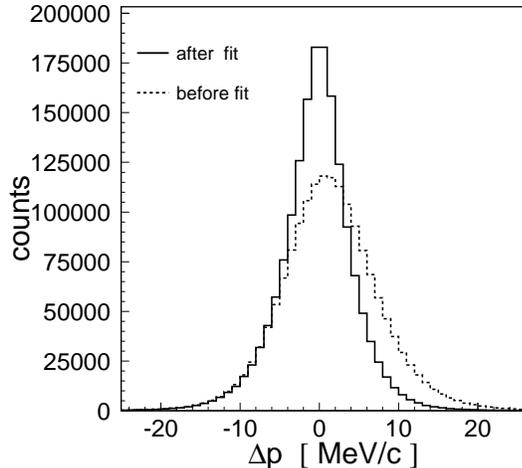}}
  \vspace{-0.5cm}
  \caption{\label{deltafit}
          Spectrum of  differences between generated
          and --~after simulation of the detector response~--
          reconstructed
          absolute momenta of  protons, 
          as determined before (dashed line) and after
          the kinematical fit (solid line). \protect\\
          The picture  shows results obtained
          taking into account the experimental conditions
          as described in the text.
  }
\end{figure}

Since the measurements have been performed close to the kinematical threshold
the ejectile momentum component parallel to the beam is by far the largest one
and
its variance ($var(p_{z})~=~37~MeV^2/c^2$)
determines in first order the error of the momentum measurement.
The second largest contribution stems from
an anti-correlation
between the $x-$ and $z-$ momentum components ($cov(p_{x},p_{z})~=~-13.7~MeV^2/c^2$),
which is due to the bending
of the proton trajectory --~mainly in the horizontal direction~--
inside the COSY-11 dipole magnet~(see fig.~\ref{detector}).
There is also a significant correlation between the $z$ components of different
protons which is due to the smearing of the reaction points,
namely, if in the analysis the assumed reaction point differs from the
actual one, a mistake made in the reconstruction affects both protons similarly.
Figure~\ref{covariance} depicts the variation of $var(p_x)$, $var(p_y)$, and $var(p_z)$
over the momentum plane ($|\vec{p_{1}}|$, $|\vec{p_{2}}|$).
Taking into account  components of the covariance matrices V($|\vec{p_{1}}|$, $|\vec{p_{2}}|$)
and the distribution of the
proton momenta for the $pp\to pp \eta$ reaction at Q~=~15.5~MeV results in an
average error for the measurement of the proton momentum of about 6~MeV/c.
This can be also deduced from the distribution of the difference between the generated
and reconstructed absolute momenta of the protons.
The corresponding spectrum is plotted as a dashed line in figure~\ref{deltafit}.

In the experiment we have measured 6 variables and once we assume that the event
corresponds to the $pp\to pp\eta$ reaction only
5 of them are independent. Thus we have varied the values of the event components
demanding that the missing mass is equal to the mass of the $\eta$ meson
and we have chosen that vector which was the closest to the experimental one.
The inverse of the covariance matrix was used as a metric for the distance
calculation.
The kinematical fit improves the  resolution
by a factor of about 1.5
as can be concluded
from comparing the dashed- and solid lines in figure~\ref{deltafit}.
 The finally resulting error of the momentum determination  amounts to 4~MeV/c.

\subsection{Multidimensional acceptance corrections and results}
\begin{figure}[b]
  \hspace{-0.9cm}\parbox{0.5\textwidth}{\vspace{-3.9cm}
    \includegraphics[width=0.57\textwidth]{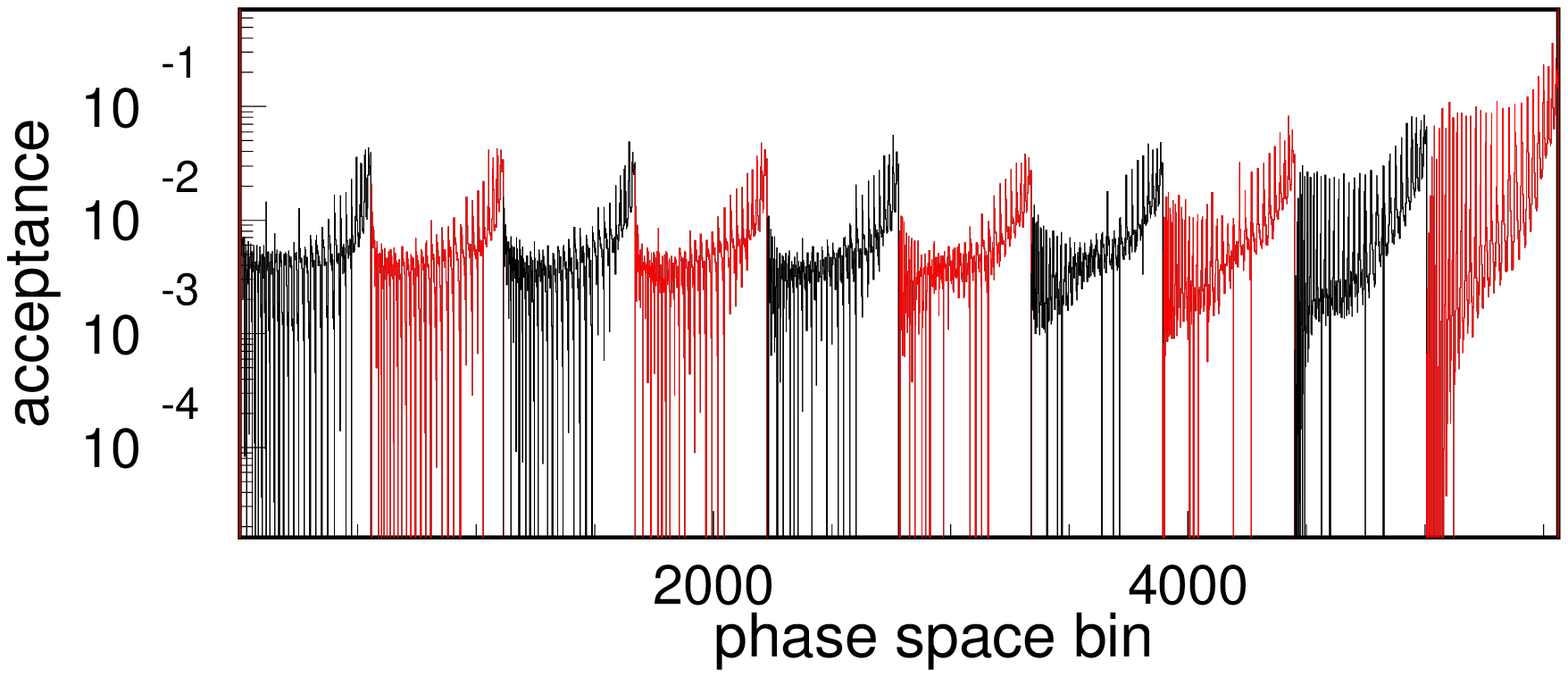}}\\
  \hspace{-0.9cm}\parbox{0.5\textwidth}{\vspace{-5.6cm}
    \includegraphics[width=0.57\textwidth]{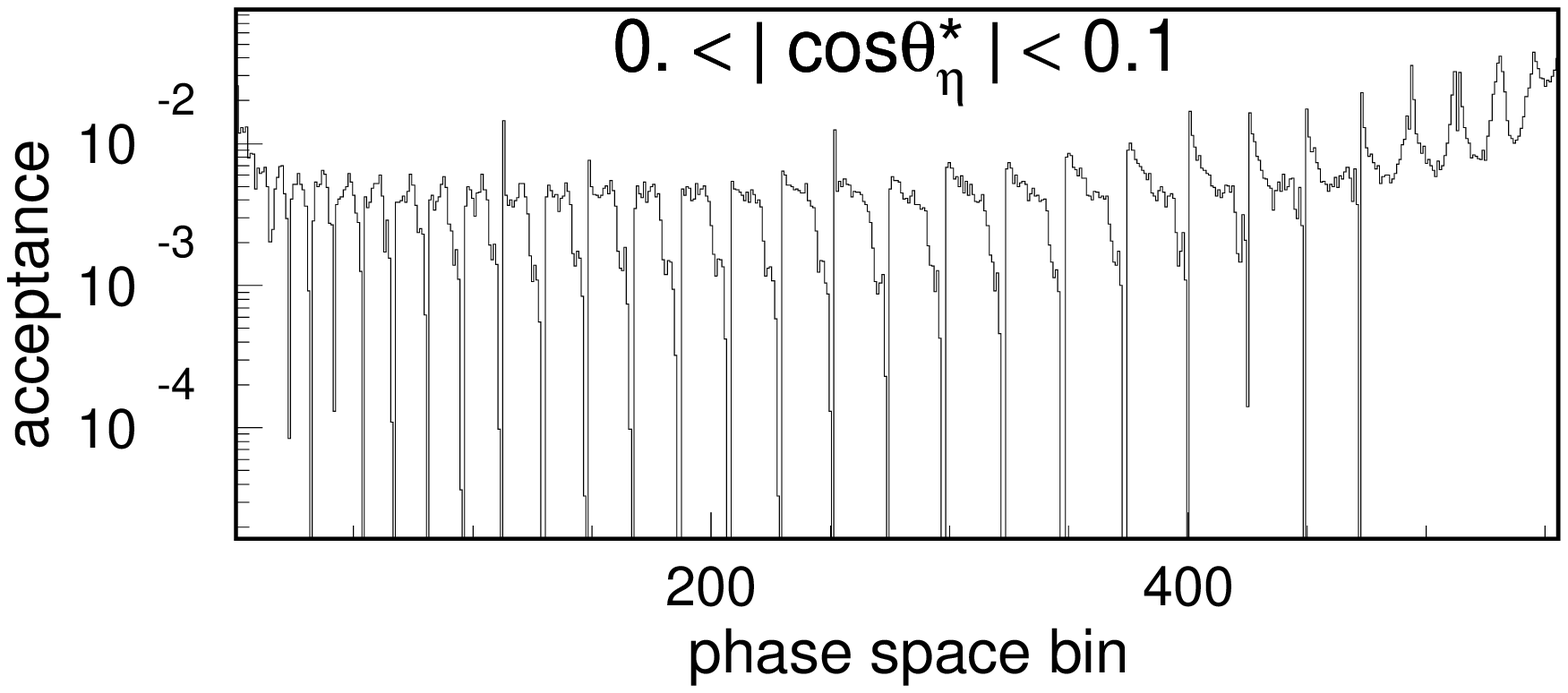}}\\
  \hspace{-0.9cm}\parbox{0.5\textwidth}{\vspace{-5.6cm}
    \includegraphics[width=0.57\textwidth]{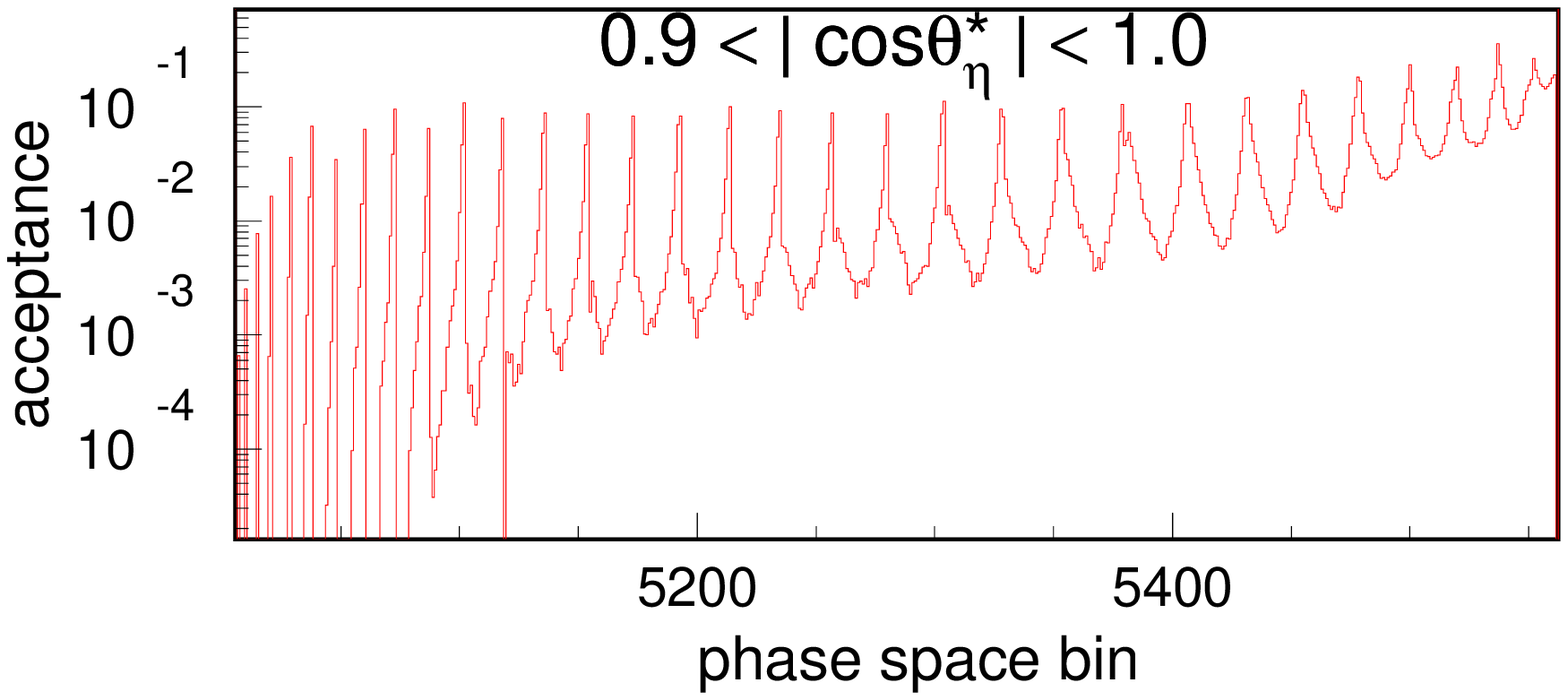}}
  \vspace{-2.5cm}
  \caption{ \label{acceptanceijk} 
    Acceptance of the COSY-11 detection system for the $pp\to pp\eta$ reaction
    at an excess energy of Q~=~15.5~MeV presented as a function of $s_{pp}$,$s_{p\eta}$,
    and $|cos(\theta_{\eta}^{*})|$.
    The numbers were assigned to the bins in the three dimensional space
    $s_{pp}-s_{p\eta}-|cos(\theta_{\eta}^{*})|$
    by first incrementing the index of $s_{pp}$ next of $s_{p\eta}$ and on the end that
    of $|cos(\theta_{\eta}^{*})|$.
    Partitioning of $|cos(\theta_{\eta}^{*})|$ 
    into ten bins 
    is easily recognizable. 
    The two lower pictures show the acceptance for the first and the last
    bin of $|cos(\theta_{\eta}^{*})|$.
  }
\end{figure}
At the excess energy of Q~=~15.5~MeV
the COSY-11 detection system does not cover the full 4$\pi$ solid angle
in the centre-of-mass  system of the $pp\to pp\eta$ reaction.
Therefore, the detailed study of differential cross sections
requires corrections for the acceptance. Generally, the acceptance
should be expressed as a function of the full set of mutually
orthogonal variables which describe the studied 
reaction unambiguously.
As introduced in section~\ref{choiceofobservables}, 
to define the relative movement of the particles in the reaction plane
we have chosen two squares of the invariant masses: $s_{pp}$  and $s_{p\eta}$,
and to define the orientation of this plane in the center-of-mass frame
we have taken the three Euler angles:
The first two are simply  the polar $\phi_{\eta}^{*}$ and azimuthal $\theta_{\eta}^{*}$ angles
of the momentum of the $\eta$ meson and the third angle $\psi$
describes the rotation of the reaction plane around the axis defined by the momentum vector
of the $\eta$ meson. 
In the data evaluation we considerably benefit from the basic geometrical symmetries
satisfied by the $pp\to pp\eta$ reaction.
 Due to the axial symmetry of  the initial channel of the two
unpolarized colliding protons the event distribution over
$\phi_{\eta}^{*}$ must be isotropic. Thus, we can safely integrate
over $\phi^{*}_{\eta}$, ignoring
that variable in the analysis.
 Furthermore, taking advantage of the symmetry
due to the two identical particles in the initial channel,
without losing the generality, we can express
the acceptance as a function of $s_{pp}$,$s_{p\eta}$,$|cos(\theta_{\eta}^{*})|$, and $\psi$.
To facilitate the calculations we have divided the range of $|cos(\theta_{\eta}^{*})|$ and $\psi$
into 10 bins and both $s_{pp}$ and $s_{p\eta}$ into 40 bins each.
In the case of the $s_{pp}$ and $s_{p\eta}$ the choice was made such
that the width of the interval corresponds to the standard deviation of the experimental
accuracy. For $|cos(\theta_{\eta}^{*})|$ and $\psi$  we have taken only ten partitions
since from the previous experiments we expect only a small variation of the cross section
over these variables~\cite{TOFeta,calen190,jim}.
In this representation, however, the COSY-11 detection system covers only 50\%
of the phase-space for the $pp\to pp\eta$ reaction at Q~=~15.5~MeV.
To proceed with the analysis we assumed that the distribution over the angle $\psi$
is isotropic as it was for example experimentally determined
for the $pp\to pp\omega$, $pp\to pp\rho$,
or $pp\to pp\phi$ reactions~\cite{balestra,jim}.
Please note that this is the only assumption
of  the reaction dynamics performed in the present evaluation.
The validity of this supposition in the case of
the $pp\to pp\eta$ reaction will be discussed later.

\begin{figure}[t]
  \parbox{0.5\textwidth}{\vspace{-1cm}
    \includegraphics[width=0.43\textwidth]{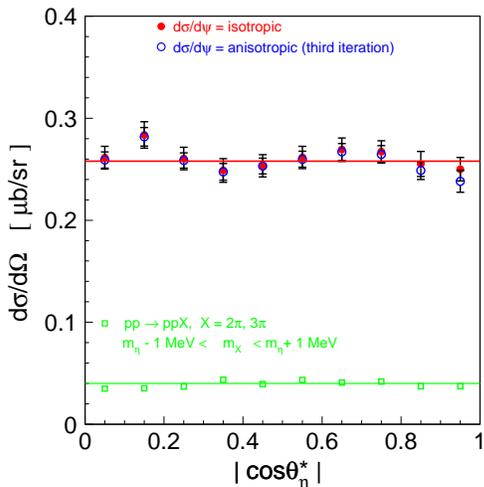}}
  \vspace{-1cm}
  \caption{ \label{rozkladthetaeta} 
    Distribution of the polar angle of the emission of the $\eta$ meson in the 
    centre--of--mass system.
    Experimental data were corrected for the acceptance
    in the three dimensional space~($s_{pp},s_{p\eta},|cos(\theta_{\eta}^{*})|)$.
    Full circles show the result with the assumption that the distribution of the $\psi$ angle
    is isotropic, and the open circles are extracted under the assumption that
    $\frac{d\sigma}{d\psi}$ is as derived from the data (see text).
    Both results have been normalized to each other in magnitude.\\
    Open squares show (in arbitrary units) the distribution of the multi-pion production
    for the invariant mass of the pions ranging between $\pm$~1~MeV around 
    the mass of the $\eta$ meson.
   }
\end{figure}
In the calculations we exploit also the symmetry of the cross sections
under the exchange of the two identical particles in the final state which reads:
$\sigma(s_{p_{1}\eta},\psi)~=~\sigma(s_{p_{2}\eta},\psi+\pi)$.
The resultant acceptance is shown in figure~\ref{acceptanceijk}.
\begin{figure}[t]
  \parbox{0.5\textwidth}{\vspace{-1.4cm}
    \includegraphics[width=0.45\textwidth]{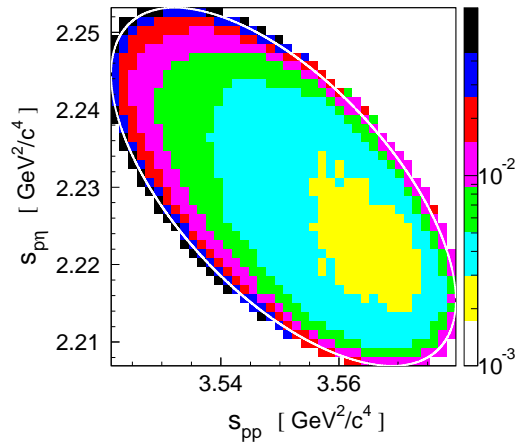}}
  \vspace{-1cm}
  \caption{ \label{acceptanceij} 
   COSY-11 detection acceptance as a function of $s_{pp}$ and $s_{p\eta}$,
   calculated under the assumption that the differential cross sections
   $\frac{d\sigma}{d\cos(\theta_{\eta}^{*})}$ and $\frac{d\sigma}{d\psi}$
   are isotropic.
   }
\end{figure}
\begin{figure}[t]
    \parbox{0.4\textwidth}{\vspace{-1.0cm} 
     \includegraphics[width=0.4\textwidth]{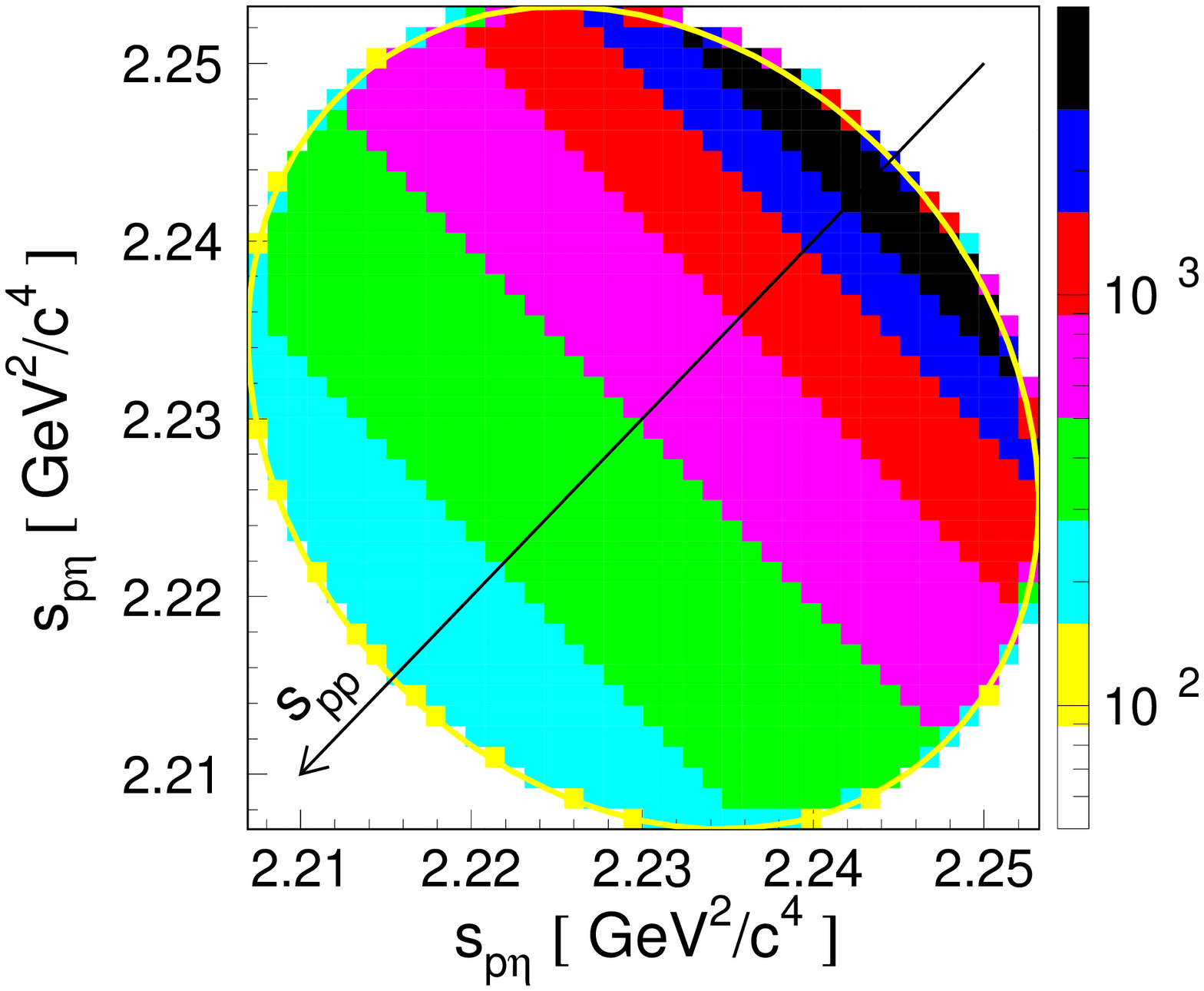}} 
    \hspace{-1cm}\parbox{0.05\textwidth}{ (a)}
    \parbox{0.4\textwidth}{\vspace{-1.4cm}
     \includegraphics[width=0.4\textwidth]{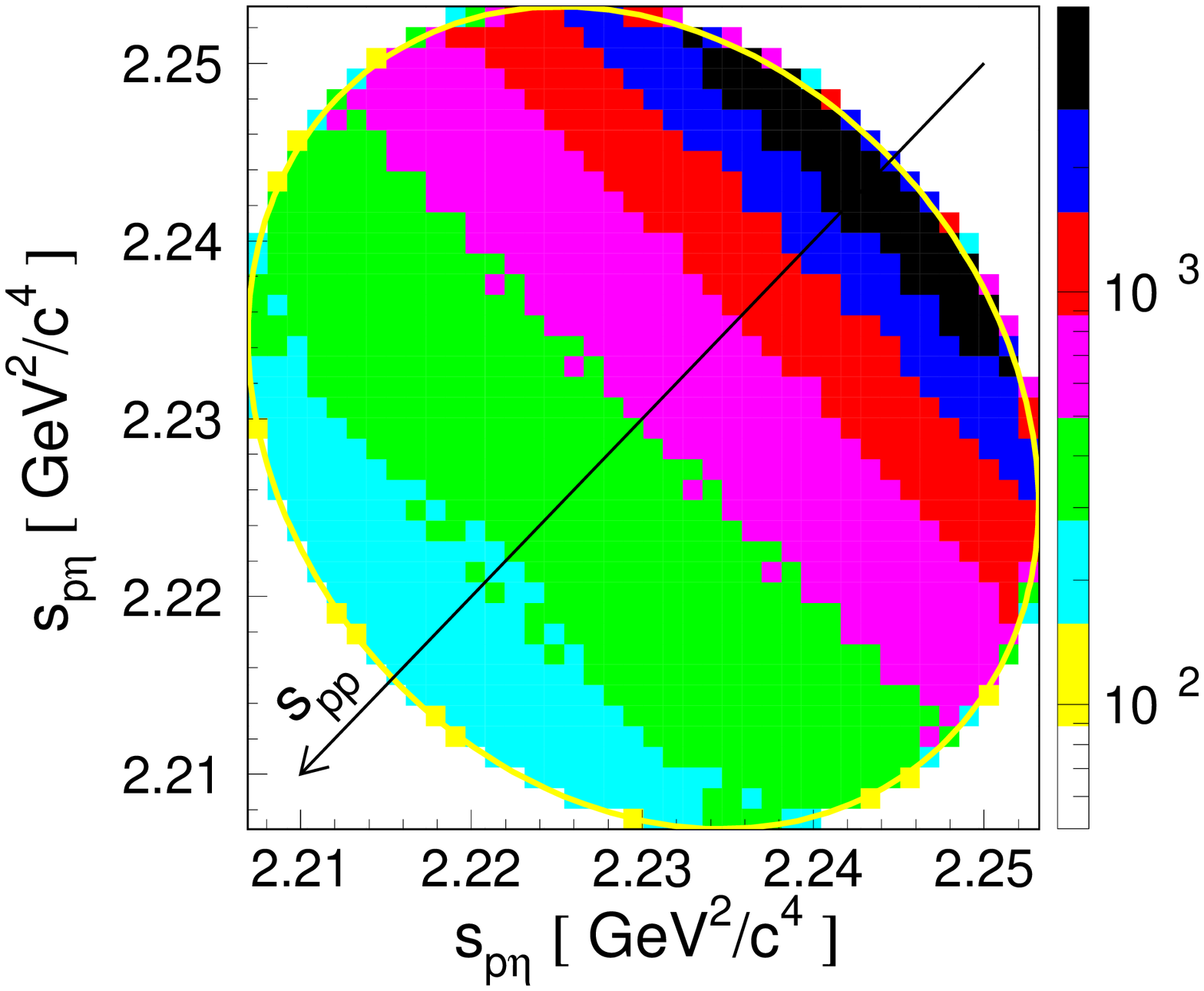}} 
    \hspace{-1cm}\parbox{0.05\textwidth}{ (b)}
  \vspace{-0.5cm}
  \caption{ \label{dalitzporownanie}
   {\bf (a)} Dalitz plot distribution simulated
    for the $pp\to pp\eta$ reaction at Q~=~15.5~MeV. In the calculations the interaction between
    protons was taken into account.\\
    {\bf (b)}
    Dalitz plot distribution reconstructed from the COSY-11 detector
    response simulated for events from figure (a) taking into account the
    smearing of the beam and target, multiple scattering in the materials,
    and the detectors resolution. The evaluation
    included momentum reconstruction, kinematical fitting and the acceptance
    correction exactly in the same way as performed for the experimental data.
    The lines surrounding the Dalitz plots depict the kinematical limits.
  } 
\end{figure}
One recognizes that only a small part (3\%) of the phase-space is not covered by the detection system.
In further calculations these  holes were corrected according to the  assumption
of a homogeneous phase space distribution. Additionally it was checked that the corrections
under other suppositions eg. regarding also the proton-proton FSI
leads to negligible differences.
Full points in figure~\ref{rozkladthetaeta} present the distribution of the polar angle of the $\eta$
meson as derived from the data after the acceptance correction. Within the statistical accuracy
it is isotropic.  Taking into account this angular distribution of the cross section
we can calculate the acceptance
as a function of $s_{pp}$ and $s_{p\eta}$ only.
 This  is shown in figure~\ref{acceptanceij}, where
one sees that now the  full phase space is covered.
This allows us to determine the distributions of $s_{pp}$ and $s_{p\eta}$.
The correctness of the performed  procedures for the simulation of the
detectors response, the  event reconstruction programs, the kinematical fitting
and acceptance correction can be confirmed by comparing
the distribution generated
(figure~\ref{dalitzporownanie}a) with the ones which underwent the complete 
analysis chain  described in this section
(figure~\ref{dalitzporownanie}b).

\begin{figure}[h]
    \parbox{0.5\textwidth}{\vspace{-1cm}
    \includegraphics[width=0.4\textwidth]{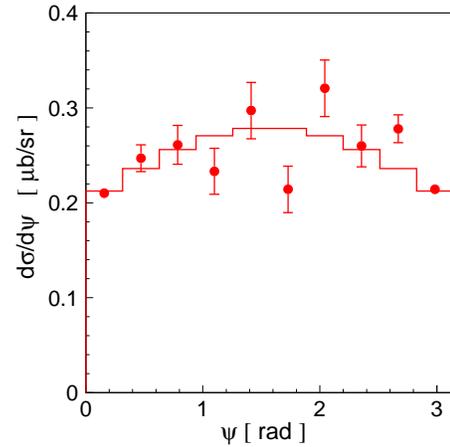}}
    \vspace{-1.0cm}
  \caption{ \label{rozkladpsi1}
       Distribution of the cross section as a function of the angle $\psi$ as determined in 
       the first iteration. The superimposed histogram corresponds to the fit 
       of the function  $\frac{d\sigma}{d\psi}~=~a~+~b~\cdot~|sin(\psi)|$.
       The range of the $\psi$ angle is shown from 0 to $\pi$ only, 
       since in the analysis we take advantage of the symmetry 
      $\frac{d\sigma}{d\psi}(\psi)~=~\frac{d\sigma}{d\psi}(\psi+\pi)$.
  } 
\end{figure}

\begin{figure}[h]
   \parbox{.5\textwidth}{ \hspace{-0.4cm}
     \vspace{-0.4cm}
     \includegraphics[width=0.26\textwidth]{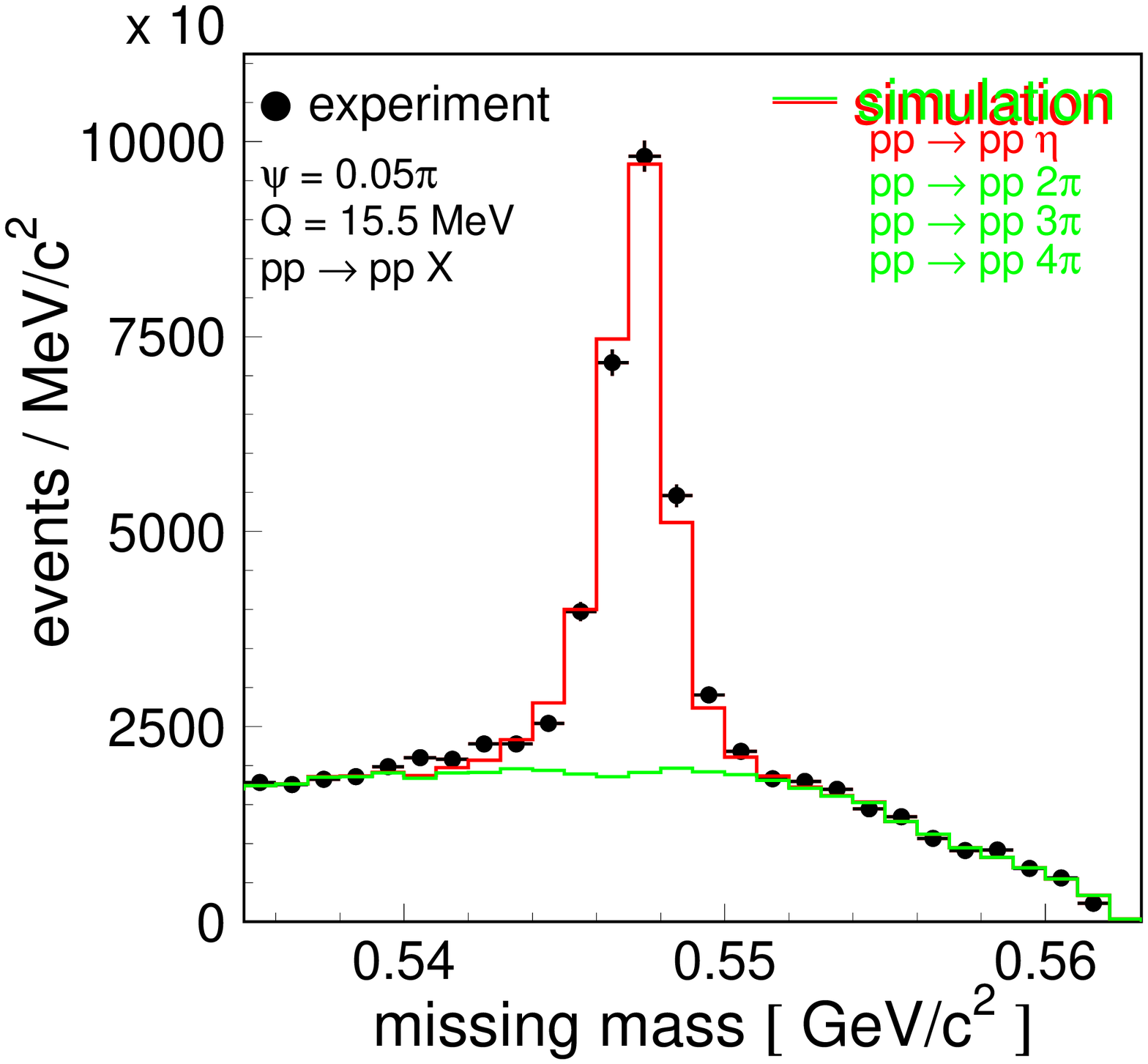}
     \hspace{-0.6cm}
     \includegraphics[width=0.26\textwidth]{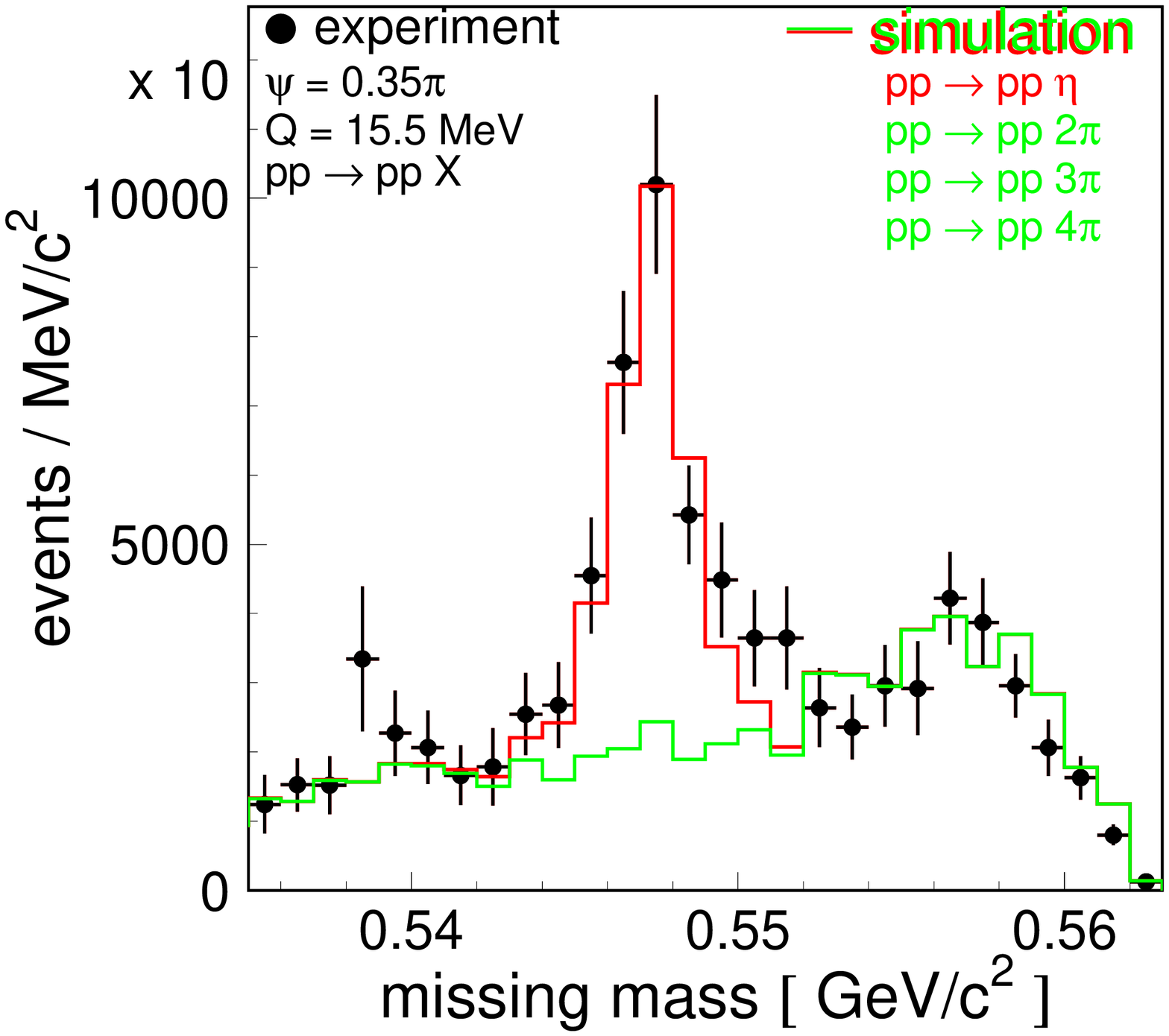}}
   \parbox{.5\textwidth}{ \hspace{-0.4cm}
     \vspace{-0.4cm}
     \includegraphics[width=0.26\textwidth]{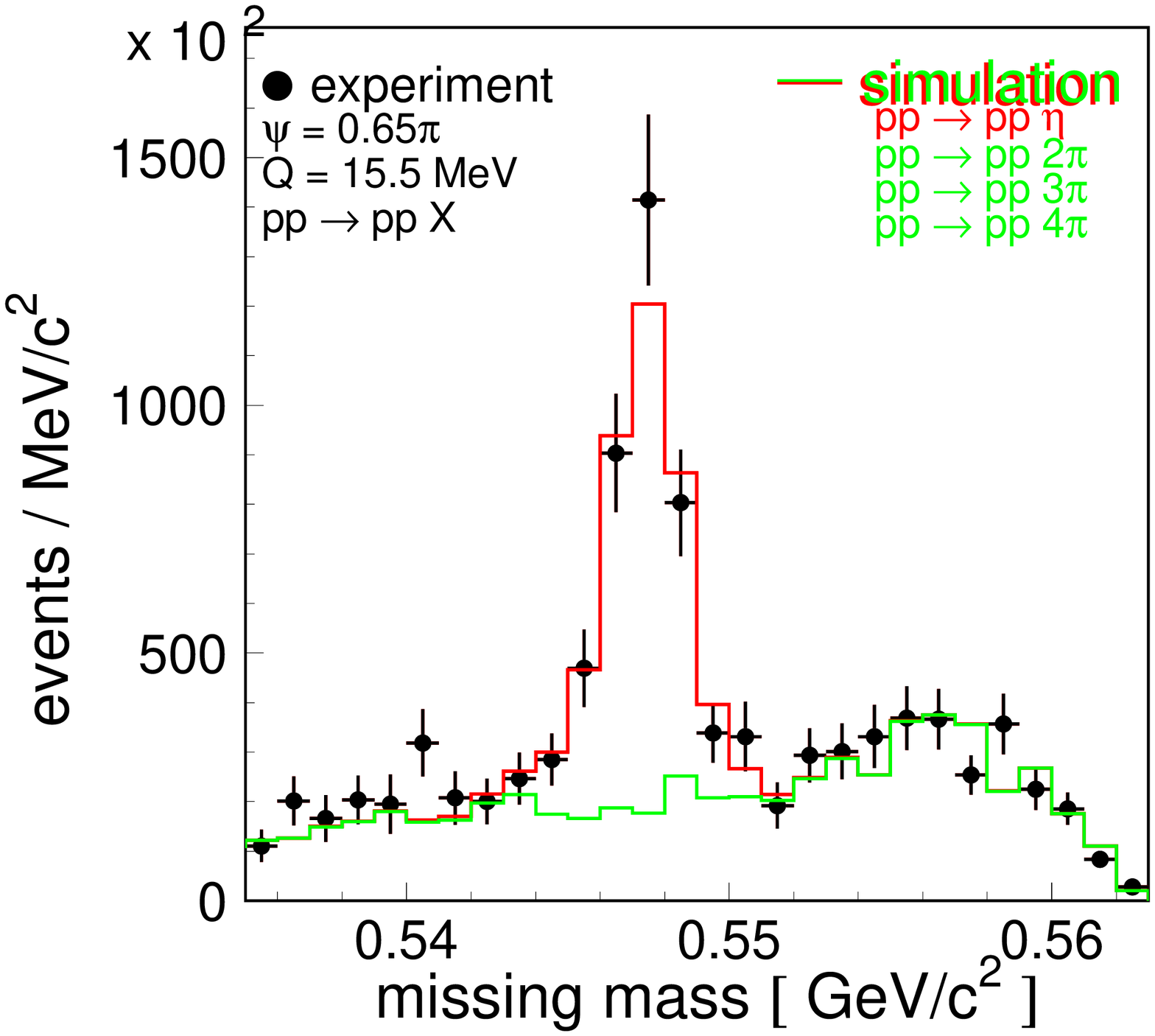}
     \hspace{-0.6cm}
     \includegraphics[width=0.26\textwidth]{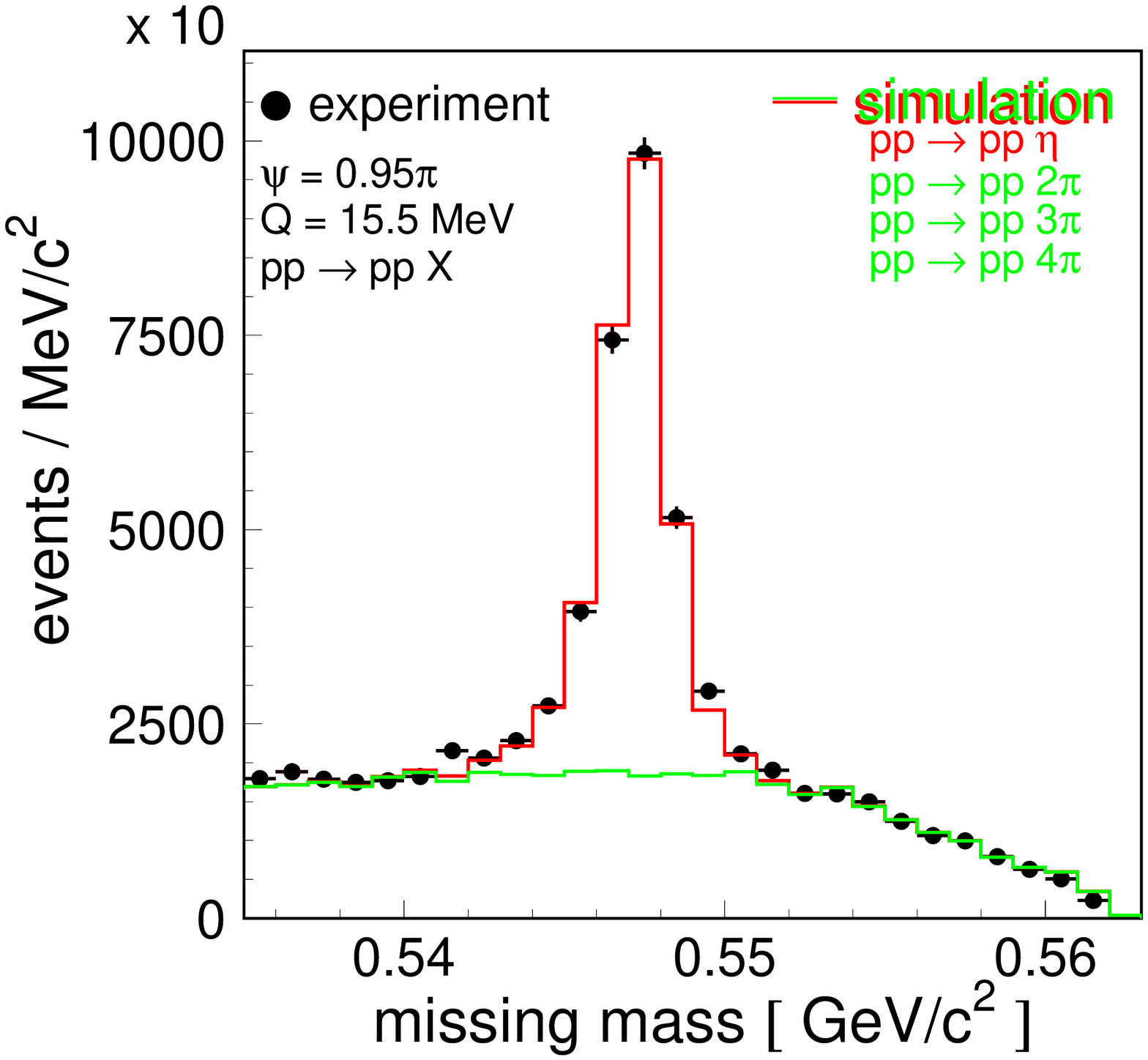}}
  \vspace{-0.2cm}
  \caption{ \label{misspsi}
     Missing mass distributions for the first, fourth, seventh and tenth  bin of $\psi$
     with the superimposed lines from the simulation 
     of $pp\to pp\eta$ and the multi-pion background $pp\to pp (m\pi)$ reactions. 
     Amplitudes of simulated distributions were fitted to the experimental points.
  }
\end{figure}

Knowing the distribution
of the polar angle of the $\eta$ meson $\theta_{\eta}^{*}$
and those for the invariant masses $s_{pp}$ and $s_{p\eta}$
we can check whether the assumption of the isotropy of the cross section
distribution versus the third Euler's angle $\psi$ is corroborated by the data.
For that purpose we calculated the acceptance as a function of
$\psi$ and $s_{p\eta}$ assuming the shape of the differential cross sections
of $\frac{d\sigma}{ds_{pp}}$ and $\frac{d\sigma}{dcos(\theta_{\eta}^{*})}$
as determined experimentally.
The obtained $\frac{d\sigma}{d\psi}$ distribution is shown
in figure~\ref{rozkladpsi1} and is not isotropic as assumed at the 
beginning.  
A fit of the  function of the form $\frac{d\sigma}{d\psi}~=~a~+~b~\cdot~|sin(\psi)|$ 
gives the value of 
 b~=~0.079~$\pm$~0.014~$\mu b / sr$, 
indeed significantly different 
from the isotropic solution.
This deviation cannot be assigned to any unknown behaviour
of the background since the
 obtained distribution can be regarded
as background free. This is because the number of $pp\to pp\eta$ events was
elaborated  for each invariant mass interval separately.
The exemplarily missing mass spectra 
for the first, fourth, seventh
and tenth interval of $\psi$ values,
corrected for the acceptance
are presented in figure~\ref{misspsi}.
\begin{figure}[t]
    \parbox{0.5\textwidth}{\vspace{-0.4cm}
    \includegraphics[width=0.33\textwidth]{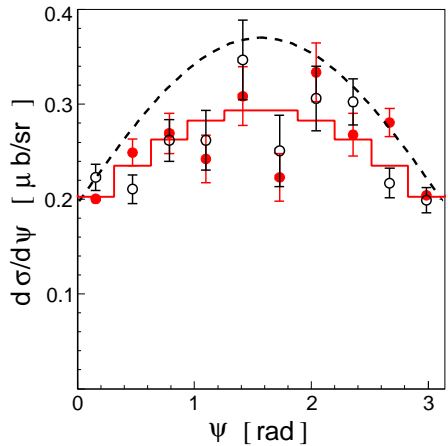}}
    \vspace{-0.5cm}
  \caption{ \label{rozkladpsi}
       Distribution of the cross section as a function 
       of the angle $\psi$.
       Full circles stand for the final results 
       of the $\frac{d\sigma}{d\psi}$
      obtained after three iterations. 
       The superimposed histogram (solid line) corresponds to the fit 
       of the function  $\frac{d\sigma}{d\psi}~=~a~+~b~\cdot~|sin(\psi)|$
       which  resulted in 
       $a~=~0.186~\pm~0.004~\mu b/sr$ and $b~=~0.110~\pm~0.014~\mu b/sr$. \\
       The dashed line shows the entry distribution used for the second series of iterations
       as described in the text.\\
     Open circles represent the 
     data from the left upper corner of the Dalitz plot 
     (see for example figure~\ref{acceptanceij}).
     At that region of the Dalitz plot due to the non-zero four dimensional acceptance
     over ($s_{pp}$, $s_{p\eta}$, $|cos(\theta_{\eta}^{*})|$,$\psi$) bins
     the spectrum (open circles) was corrected
     without a necessity of any assumptions concerning the reaction cross section.
  } 
\end{figure}

From this figure one can infer that the shape of the background is well
reproduced not only for the overall missing mass spectrum as shown
previously in figure~\ref{missall} but also locally  in each
region of the phase space.
Since the experimental data are quite well described by the simulations 
we can 
rather exclude the possibility of a significant systematical error
which could cause the observed anisotropy 
of the differential cross section $\frac{d\sigma}{d\psi}$.

The evaluated distribution is however  
in disagreement with our working assumption that the $\sigma(\psi)$ is isotropic.
Therefore we performed a full acceptance correction procedure 
from the very beginning assuming that the distribution of $\frac{d\sigma}{d\psi}$ is as 
determined from the data. After repeating the procedure three times
we observed that the input and resultant  distributions are in good agreement. 
The result after the third iteration is shown in figure~\ref{rozkladpsi} by the full circles.
It is only slightly different from the one obtained after the first iteration
as shown in figure~\ref{rozkladpsi1}.
To rise the confidence  of the convergence of the performed iteration
we accomplished  the full procedure once more, but now assuming that the distribution 
of $\frac{d\sigma}{d\psi}$ is much more anisotropic 
than determined from the data. As an entry distribution we took the 
dashed line shown in figure~\ref{rozkladpsi}.
Again after two iterations we have got the same result as before.
Finally determined values for  $\frac{d\sigma}{d\psi}(\psi)$ are given 
in table~\ref{tablepsi}.
\begin{table}[t]
\caption{\label{tablepsi}
  Differential cross section in the $\psi$ angle for the $pp\to pp\eta$ reaction 
  at Q~=~15.5~MeV.
  Fitting to the data a function of the form
  $\frac{d\sigma}{d\psi}~=~a~+~b~|sin(\psi)|$ resulted in 
  $a~=~0.186~\pm~0.004~\mu b/sr$ and $b~=~0.110~\pm~0.014~\mu b/sr$.
  Figure~\ref{rozkladpsi} illustrates the result.
}
\begin{ruledtabular}
\begin{tabular}{lcr}
$\displaystyle{\psi [rad]}$ 
& $\displaystyle{\frac{d\sigma}{d\psi}}$ $\displaystyle{\left[\frac{\mu b}{rad}\right]}$ 
& $\displaystyle{\frac{d\sigma}{d\psi}}$ $\displaystyle{\left[\frac{\mu b}{rad}\right]}$ \\
 & experiment &  fit \\
\hline
0.157 &0.200~$\pm$~0.004 &0.202 \\
0.471 &0.249~$\pm$~0.014 &0.235 \\
0.785 &0.269~$\pm$~0.021 &0.263 \\
1.100 &0.242~$\pm$~0.025 &0.283 \\
1.414 &0.309~$\pm$~0.031 &0.294 \\
1.728 &0.223~$\pm$~0.025 &0.294 \\
2.042 &0.334~$\pm$~0.031 &0.283 \\
2.356 &0.268~$\pm$~0.023 &0.263 \\
2.670 &0.281~$\pm$~0.015 &0.235 \\
2.985 &0.204~$\pm$~0.004 &0.202 \\
\end{tabular}
\end{ruledtabular}
\end{table}
To corroborate this observation
we have evaluated the distribution over $\psi$ angle (see figure~\ref{rozkladpsi})
from the phase space region which
has no holes in the acceptance expressed as a four-dimensional function of
the variables $s_{pp}$, $s_{p\eta}$, $|cos(\theta_{\eta}^{*})|$, and $\psi$,
this is for the values of $s_{pp}$ and $s_{p\eta}$ corresponding to the 
upper left corner of figure~\ref{acceptanceij}.
 Again the obtained distribution 
presented as open circles in figure~\ref{rozkladpsi} is anisotropic,
and moreover agrees
with the spectrum determined from all events.
The anisotropy of the cross section in the $\psi$ angle 
reflects itself in an anisotropy of the orientation of the emission plane.

\begin{figure}[h]
\parbox{0.5\textwidth}{
 \vspace{-0.4cm}
 \includegraphics[width=0.4\textwidth]{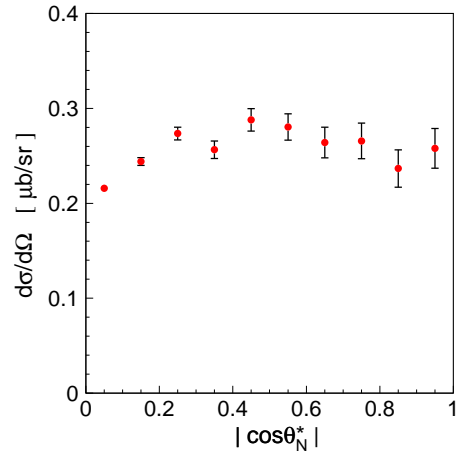}}
 \vspace{-0.7cm}
 \caption{\label{thetaN} 
  Differential cross section as a function of 
  the polar angle of the vector normal to the emission plane.}
\end{figure}

The determined cross section distribution in function 
of the polar angle $\theta^*_{N}$ of the vector normal to that plane
is shown in figure~\ref{thetaN} and the corresponding values are 
listed in table~\ref{tablecosinusy}. 
The distribution is not isotropic,
which is particularly visible for the low values of $|cos(\theta^*_{N})|$
burdened with  small errors.
As depicted in figure~\ref{figvariables} the $|cos(\theta^*_{N})|$~=~0
denotes such configuration of the ejectiles momenta in which the emission plane
comprises the beam axis. In that case the acceptance of the COSY-11 detection
system is much larger than for the configuration where the 
emission plane  is perpendicular to the beam.
Due to this reason the error bars  in figure~\ref{thetaN}
increase with growth of  $|cos(\theta^*_{N})|$.
It is worth to stress that the tendency 
of the $pp\eta$ system  to be produced 
preferentially
if the emission plane is perpendicular to the beam 
is in line with the preliminary analysis of the experiment performed 
by the TOF collaboration~\cite{Eduardacta}. Elucidation of that non-trivial 
behaviour can reveal interesting features of the dynamics of the 
production process. 

\begin{figure}[t]
  \parbox{0.5\textwidth}{\vspace{-0.4cm}
    \includegraphics[width=0.4\textwidth]{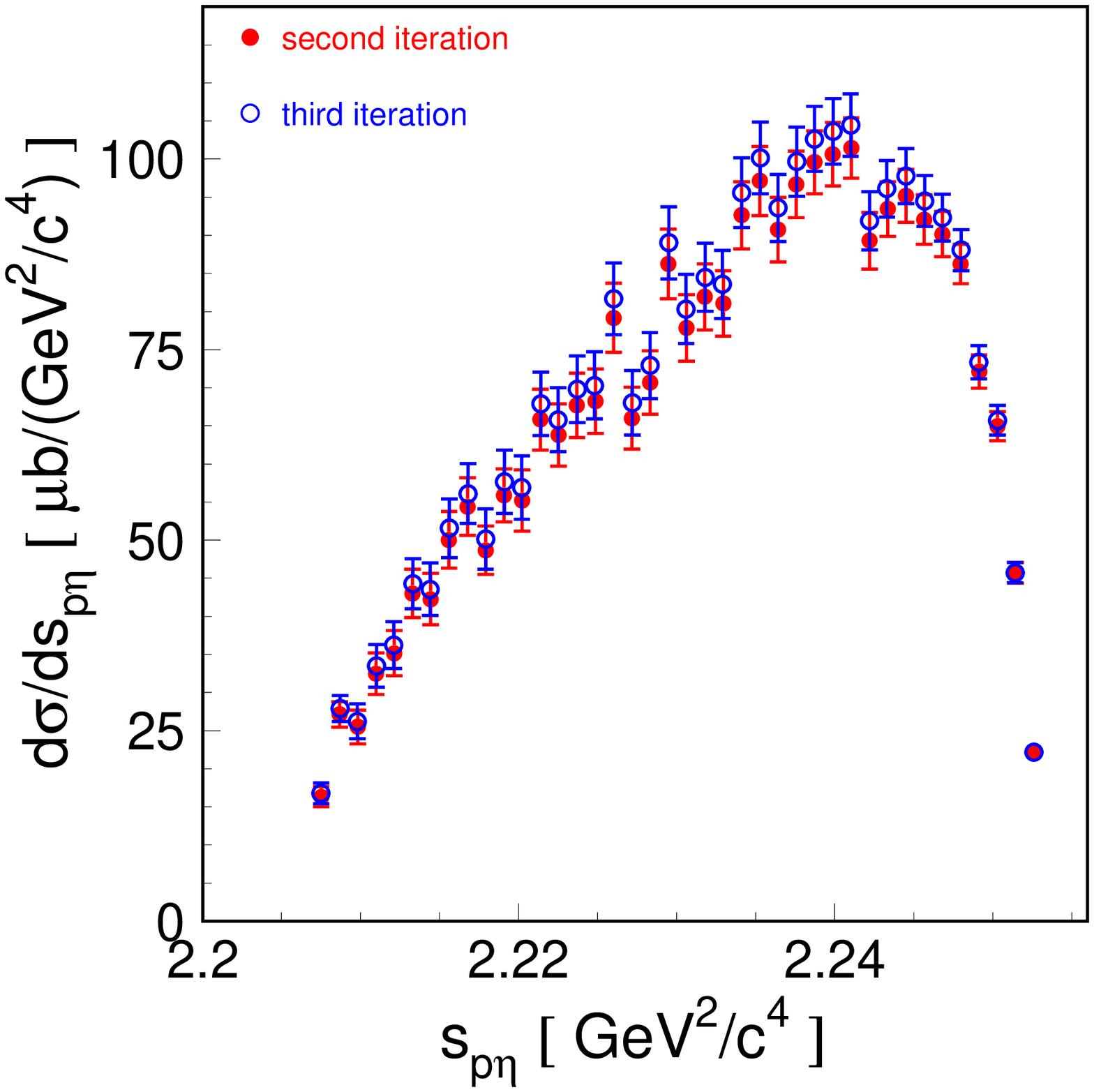}}
  \parbox{0.5\textwidth}{\vspace{-0.5cm}
    \includegraphics[width=0.4\textwidth]{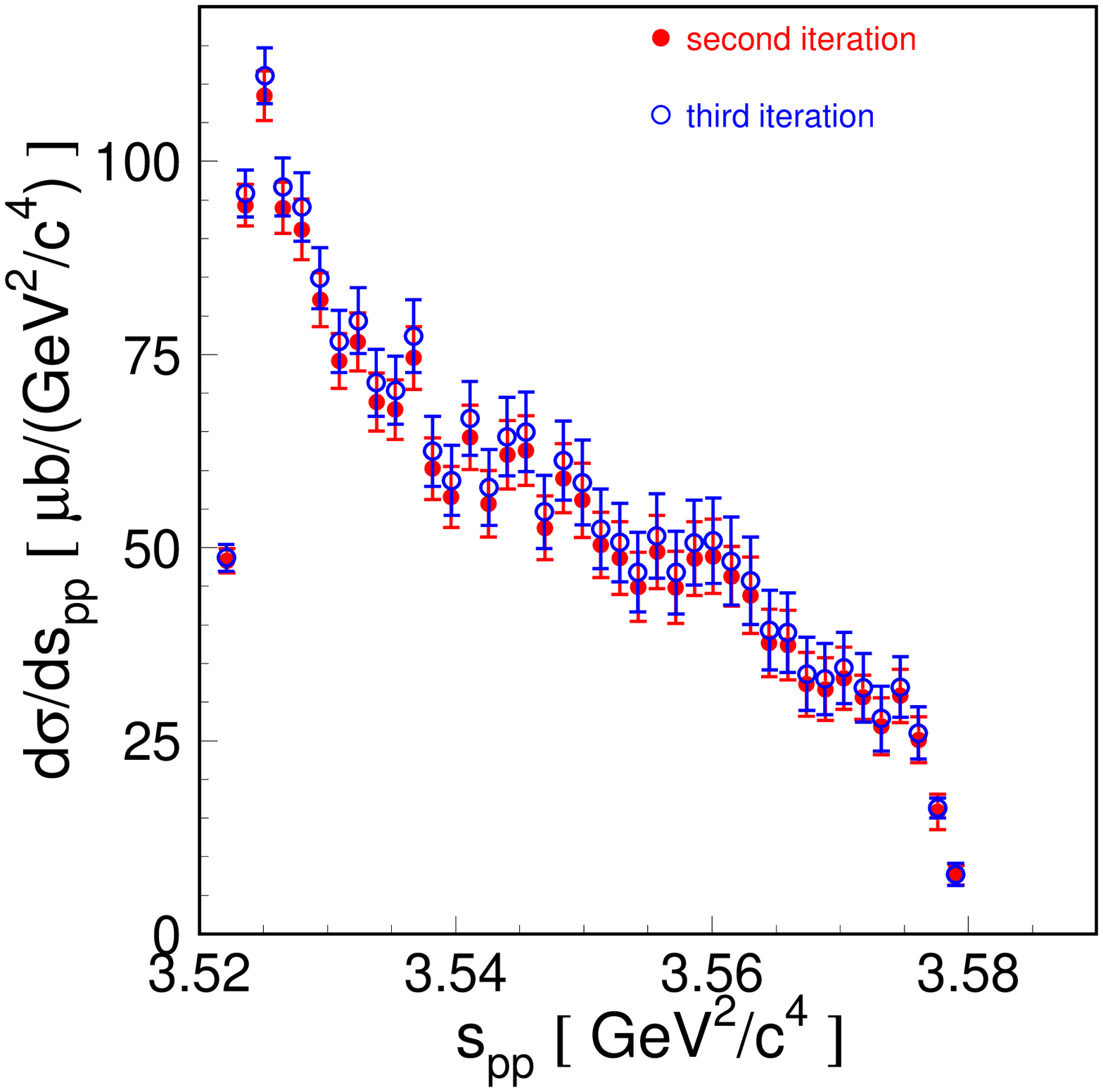}}
  \vspace{-0.6cm}
  \caption{ \label{invporownanie} 
      Distributions of the invariant masses $s_{pp}$ and $s_{p\eta}$ determined for the
      two different assumptions about the cross section dependence of the $\psi$ angle.
  }
\end{figure}

It is important to note that the shape of the $s_{pp}$, $s_{p\eta}$, and 
$cos(\theta_{\eta}^{*})$ distributions keeps unchanged during the whole 
iteration procedure.  Figure~\ref{invporownanie}
shows the distributions of the square of the proton-proton and proton-$\eta$
invariant masses. The spectra after the second and third
iterations are shown. One recognizes that the form of the spectra remains unaltered.
The same 
conclusion can be drawn for the $\frac{d\sigma}{d|cos(\theta^{*}_\eta)|}$ distribution
as demonstrated in figure~\ref{rozkladthetaeta}. 
From that comparison one can conclude
that the shapes of the determined distributions are
--~within the statistical accuracy~-- independent
of the shape of the $\frac{d\sigma}{d\psi}$ cross section
and can be treated as derived in a completely model independent manner.
Similarly, as in the case of the $\frac{d\sigma}{d\psi}$ distribution,
the differential cross sections 
in all other variables reported in this article
are not deteriorated  by the background. This is because the 
number of $pp\to pp\eta$ events was
determined  for each investigated  phase-space interval separately.
\begin{figure}[t]
    \parbox{0.5\textwidth}{
     \vspace{-0.9cm}
     \includegraphics[width=0.35\textwidth]{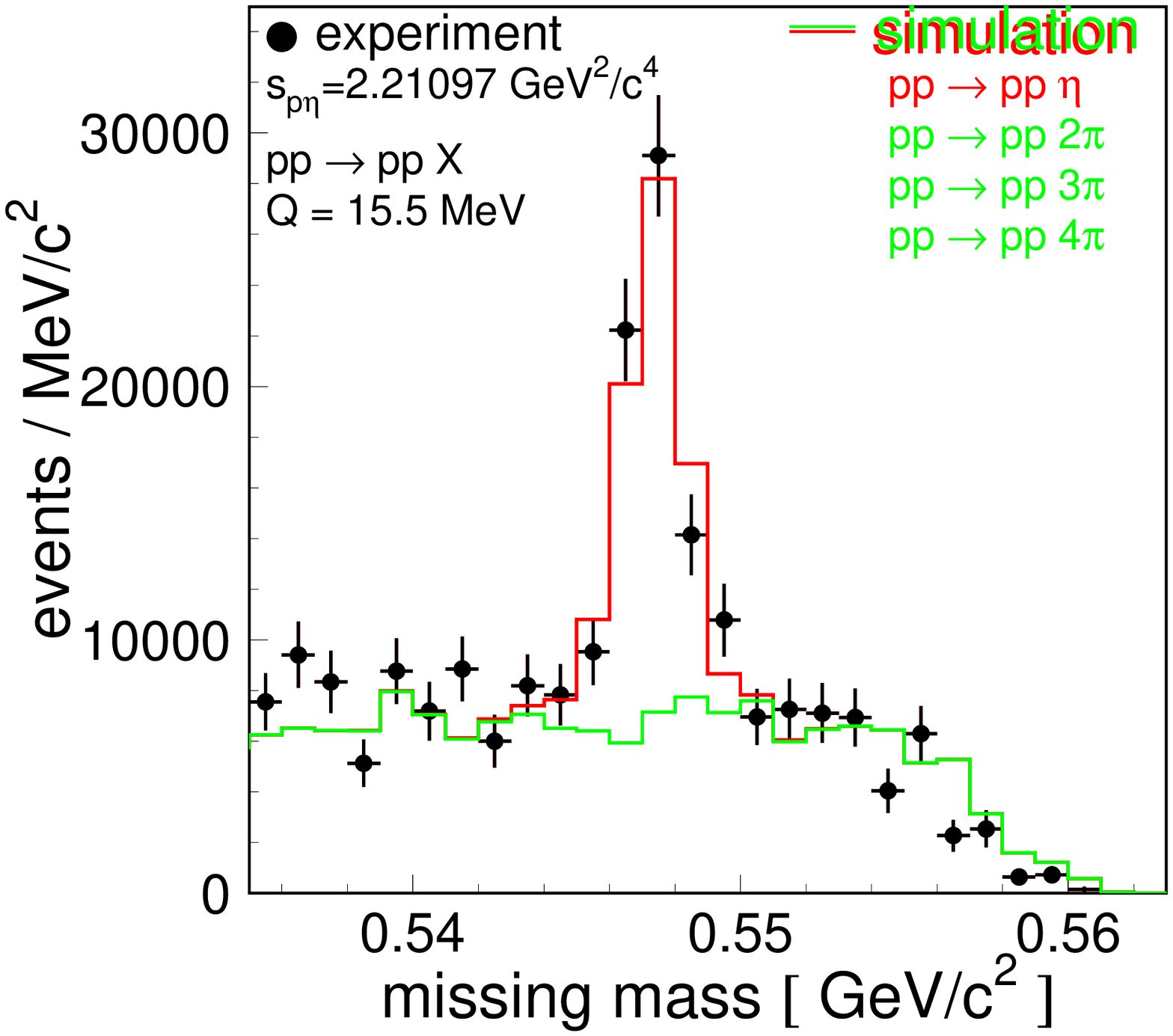}
     
     \vspace{-0.9cm}
     \includegraphics[width=0.35\textwidth]{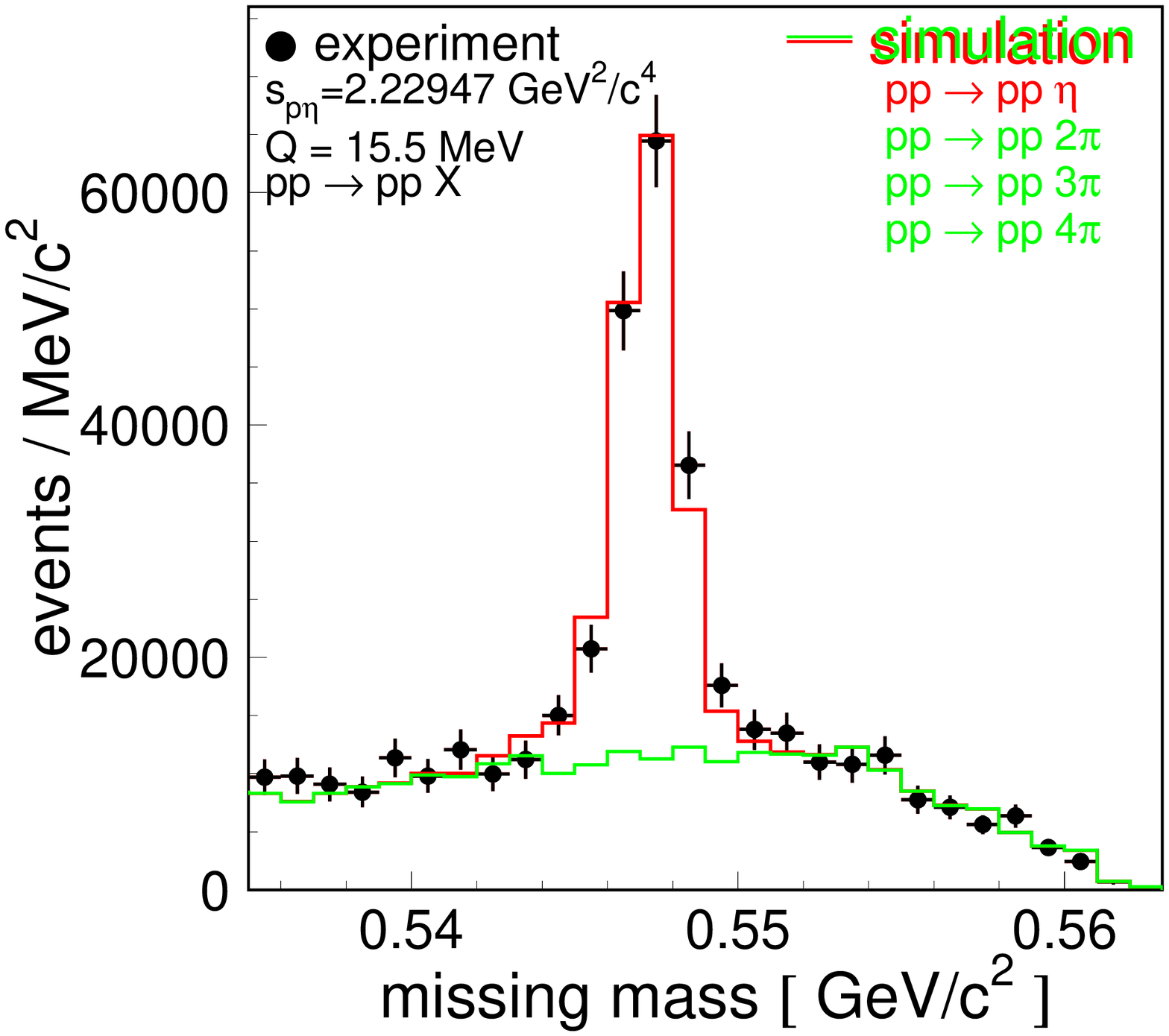}
     
     \vspace{-0.9cm}
     \includegraphics[width=0.35\textwidth]{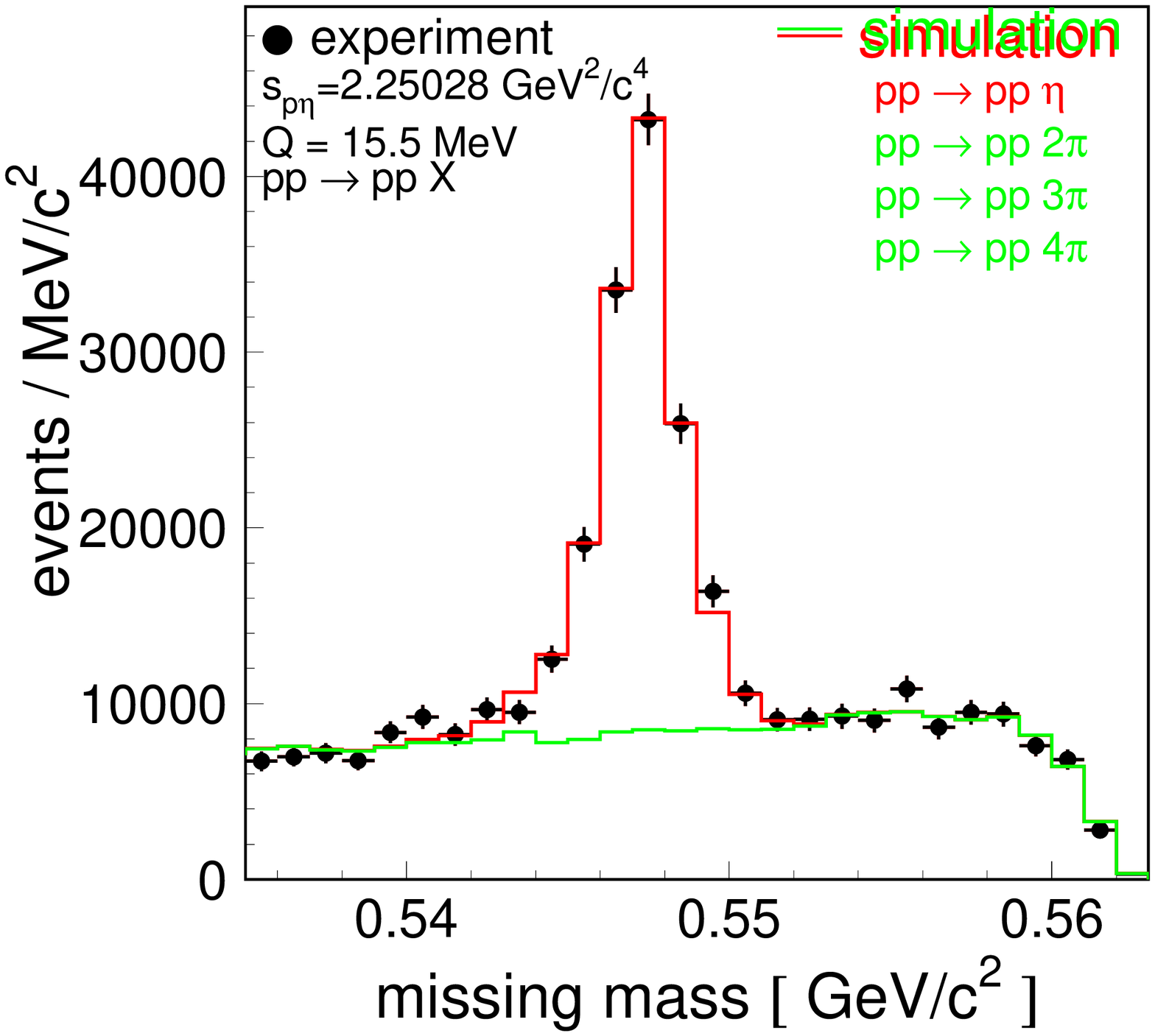}}
   \vspace{-0.7cm}
  \caption{ \label{invmiss}
     Missing mass distributions determined for the invariant mass bins as depicted inside
     the figures.  The spectra were corrected for the acceptance.
     The  histograms show the simulations for the
     multi-pion background  $pp\to pp (m\pi)$ and the $pp\to pp \eta$ reactions 
     fitted to the data with the amplitudes
     as the only free parameters.
  }
\end{figure}
As an example 
the missing mass spectra 
for three bins of the proton-$\eta$ invariant mass 
are presented in figure~\ref{invmiss}. 
As already noticed for $\frac{d\sigma}{d\psi}$ 
the shape of the background is well
reproduced locally  in each region of the phase space.

\subsection{Total and differential cross sections}
Though the form of the $s_{pp}$, $s_{p\eta}$, and $cos(\theta^{*}_{\eta})$ 
is independent 
of the $\psi$ distribution,
the total cross section derived from the data 
depends on the shape of  $\frac{d\sigma}{d\psi}$  quite significantly.
It amounts to 3.24~$\pm$~0.03~$\mu b$,
yet it changes by $\pm$~0.2~$\mu b$ when  varying  the
parameters of the function $\frac{d\sigma}{d\psi}~=~a~+~b~\cdot~|sin(\psi)|$
by $\pm$  three standard deviations. 
Therefore, we use that variation as an estimation of the systematic
error in the acceptance correction. To this we must add a $3\%$ systematical 
uncertainty stemming  from the luminosity determination~\cite{phd}. 
The luminosity was determined by the comparison of the measured differential
distribution of the elastically scattered protons with the results of the 
EDDA collaboration~\cite{EDDA}. The determined value amounts to $811\pm 8~\pm (3\%)$~nb$^{-1}$.
Thus the overall
systematical error of the cross section value amounts to 0.30~$\mu b$. 
In summary, we determined that at 
Q~=~15.5~MeV the total cross section for the $pp\to pp\eta$ reaction 
is equal to 3.24~$\pm$~0.03~$\pm$~0.30~$\mu b$, where the first and second error
denote the statistical and systematical uncertainty, 
respectively~\cite{TOTAL}.

\begin{figure}[t]
    \parbox{0.5\textwidth}{\vspace{-0.4cm}
    \includegraphics[width=0.4\textwidth]{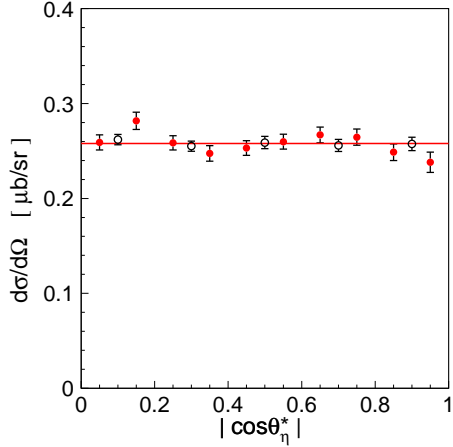}}
    \vspace{-0.9cm}
    \caption{\label{costhetaetatof}
     Differential cross section of the $pp\to pp\eta$ reaction as a function of the
     $\eta$ meson centre-of-mass polar angle. \\
     Full circles depict experimental results
     for the $pp\to pp\eta$ reaction measured at Q~=~15.5~MeV by the COSY-11 collaboration
     (this article) and the open circles were determined by the TOF collaboration
     at Q~=~15~MeV~\cite{TOFeta}. The TOF points were normalized in amplitude to our result,
     since for that data the absolute scale is not evaluated.
    }
   \end{figure}

In the below tables and figures we will present the values for the 
differential cross section 
concerning the observables 
described in section~\ref{choiceofobservables}.
  If possible the data will be compared to the 
result of measurements performed at the non-magnetic spectrometer COSY-TOF~\cite{TOFeta}.
An interpretation of the elaborated distribution follows in the next section.
The distributions  of the squared invariant masses
as presented already in figure~\ref{invporownanie},
are listed in table~\ref{tablesppspeta}. The distribution of the polar angle 
of the $\eta$ meson  emission in the centre-of-mass system is given 
in table~\ref{tablecosinusy} and shown in figure~\ref{costhetaetatof}.
\begin{table}[t]
\caption{\label{tablecosinusy}
 Differential cross section in $|cos(\theta_{\eta}^{*})|$,
 $|cos(\theta_{N}^{*})|$, $|cos(\theta_{pp}^{*})|$, and $|cos(\theta_{pp}^{**})|$
 for the $pp\to pp\eta$ reaction at Q~=~15.5~MeV.
 }
\begin{ruledtabular}
\begin{tabular}{l|c|c|c|c}
&$\frac{d\sigma}{d\Omega}(|cos(\theta_{\eta}^{*})|)$    
&$\frac{d\sigma}{d\Omega}(|cos(\theta_{N}^{*})|)$    
&$\frac{d\sigma}{d\Omega}(|cos(\theta_{pp}^{*})|)$    
&$\frac{d\sigma}{d\Omega}(|cos(\theta_{pp}^{**})|)$ \\ 
 & [$\frac{\mu b}{sr}$] & [$\frac{\mu b}{sr}$] 
 & [$\frac{\mu b}{sr}$] & [$\frac{\mu b}{sr}$] \\
\hline
.05 &.259~$\pm$~.008 &.216~$\pm$~.003 &.256~$\pm$~.010 &.276~$\pm$~.010 \\
.15 &.282~$\pm$~.009 &.244~$\pm$~.004 &.259~$\pm$~.010 &.267~$\pm$~.010 \\
.25 &.259~$\pm$~.007 &.273~$\pm$~.007 &.279~$\pm$~.010 &.266~$\pm$~.009 \\
.35 &.248~$\pm$~.008 &.256~$\pm$~.009 &.266~$\pm$~.009 &.257~$\pm$~.009 \\
.45 &.253~$\pm$~.008 &.288~$\pm$~.012 &.251~$\pm$~.010 &.255~$\pm$~.009 \\
.55 &.260~$\pm$~.008 &.280~$\pm$~.014 &.268~$\pm$~.009 &.273~$\pm$~.008 \\
.65 &.267~$\pm$~.008 &.264~$\pm$~.016 &.260~$\pm$~.009 &.252~$\pm$~.008 \\
.75 &.265~$\pm$~.009 &.266~$\pm$~.019 &.253~$\pm$~.007 &.247~$\pm$~.007 \\
.85 &.249~$\pm$~.009 &.237~$\pm$~.020 &.245~$\pm$~.006 &.259~$\pm$~.006 \\
.95 &.238~$\pm$~.011 &.258~$\pm$~.021 &.240~$\pm$~.003 &.229~$\pm$~.004 \\
\end{tabular}
\end{ruledtabular}
\end{table}
Clearly, our data agree very well with the angular dependence determined 
by the TOF collaboration.
As already mentioned, the two identical particles in the initial state of the $pp\to pp\eta$ reaction
imply that the angular distribution of either ejectile must be symmetric around
90 degree in the centre-of-mass frame. 
Here we use that reaction characteristic, yet
in the previous analysis of this data \cite{menu2001}
we utilized this property to 
check the correctness of the acceptance calculation.
In reference~\cite{menu2001} we presented the differential cross section
of the $\eta$ meson centre-of-mass polar angle for the full range of $cos(\theta^{*}_{\eta})$
and found that this is 
completely symmetric around $cos(\theta^{*}_{\eta})~=~0$.
\begin{table}[b]
\caption{\label{tablesppspeta} Distribution of the square of the invariant mass 
   of the proton-proton  ($s_{pp}$) and proton-$\eta$ ($s_{p\eta}$) systems 
   measured at Q~=~15.5~MeV via $pp\to pp\eta$ reaction.
}
\begin{ruledtabular}
\begin{tabular}{cc||cc}
$\displaystyle{s_{pp}}     $ &  $\displaystyle{\frac{d\sigma}{ds_{pp}}}$    &
$\displaystyle{s_{p\eta}}  $ &  $\displaystyle{\frac{d\sigma}{ds_{p\eta}}}$ \\
$\displaystyle{[GeV^2/c^4]}$ &  $\displaystyle{\left[\frac{\mu b}{GeV^2/c^4}\right]}$ &
$\displaystyle{[GeV^2/c^4]}$ &  $\displaystyle{\left[\frac{\mu b}{GeV^2/c^4}\right]}$ \\
\hline
3.5221 & 48.7~$\pm$~1.7 &2.2075 & 16.8~$\pm$~1.4 \\
3.5236 & 95.8~$\pm$~3.0 &2.2087 & 27.9~$\pm$~1.7 \\
3.5251 &111.1~$\pm$~3.6 &2.2098 & 26.2~$\pm$~2.3 \\
3.5265 & 96.7~$\pm$~3.8 &2.2110 & 33.5~$\pm$~2.8 \\
3.5280 & 94.1~$\pm$~4.4 &2.2121 & 36.2~$\pm$~3.1 \\
3.5294 & 84.9~$\pm$~4.0 &2.2133 & 44.3~$\pm$~3.3 \\
3.5309 & 76.7~$\pm$~4.0 &2.2144 & 43.6~$\pm$~3.5 \\
3.5324 & 79.4~$\pm$~4.3 &2.2156 & 51.6~$\pm$~3.8 \\
3.5338 & 71.3~$\pm$~4.3 &2.2168 & 56.1~$\pm$~3.9 \\
3.5353 & 70.4~$\pm$~4.4 &2.2179 & 50.2~$\pm$~4.0 \\
\hline
3.5367 & 77.4~$\pm$~4.7 &2.2191 & 57.6~$\pm$~4.2 \\
3.5382 & 62.5~$\pm$~4.5 &2.2202 & 56.9~$\pm$~4.2 \\
3.5397 & 58.7~$\pm$~4.5 &2.2214 & 67.9~$\pm$~4.2 \\
3.5411 & 66.7~$\pm$~4.8 &2.2225 & 65.8~$\pm$~4.2 \\
3.5426 & 57.8~$\pm$~4.9 &2.2237 & 69.8~$\pm$~4.4 \\
3.5440 & 64.4~$\pm$~5.1 &2.2248 & 70.3~$\pm$~4.4 \\
3.5455 & 65.0~$\pm$~5.1 &2.2260 & 81.7~$\pm$~4.7 \\
3.5469 & 54.6~$\pm$~4.7 &2.2272 & 68.0~$\pm$~4.2 \\
3.5484 & 61.3~$\pm$~5.1 &2.2283 & 72.9~$\pm$~4.3 \\
3.5499 & 58.4~$\pm$~5.5 &2.2295 & 89.0~$\pm$~4.7 \\
\hline
3.5513 & 52.4~$\pm$~5.2 &2.2306 & 80.3~$\pm$~4.5 \\
3.5528 & 50.7~$\pm$~5.1 &2.2318 & 84.5~$\pm$~4.5 \\
3.5542 & 46.8~$\pm$~5.1 &2.2329 & 83.6~$\pm$~4.4 \\
3.5557 & 51.5~$\pm$~5.4 &2.2341 & 95.6~$\pm$~4.6 \\
3.5572 & 46.8~$\pm$~5.3 &2.2353 &100.2~$\pm$~4.7 \\
3.5586 & 50.7~$\pm$~5.5 &2.2364 & 93.6~$\pm$~4.4 \\
3.5601 & 50.9~$\pm$~5.5 &2.2376 & 99.7~$\pm$~4.5 \\
3.5615 & 48.3~$\pm$~5.7 &2.2387 &102.6~$\pm$~4.3 \\
3.5630 & 45.7~$\pm$~5.7 &2.2399 &103.6~$\pm$~4.3 \\
3.5645 & 39.3~$\pm$~5.2 &2.2410 &104.5~$\pm$~4.1 \\
\hline
3.5659 & 39.0~$\pm$~5.2 &2.2422 & 91.9~$\pm$~3.8 \\
3.5674 & 33.6~$\pm$~4.7 &2.2433 & 96.1~$\pm$~3.7 \\
3.5688 & 33.0~$\pm$~4.6 &2.2445 & 97.8~$\pm$~3.6 \\
3.5703 & 34.4~$\pm$~4.6 &2.2457 & 94.5~$\pm$~3.4 \\
3.5718 & 31.8~$\pm$~4.4 &2.2468 & 92.3~$\pm$~3.0 \\
3.5732 & 27.9~$\pm$~4.2 &2.2480 & 88.1~$\pm$~2.7 \\
3.5747 & 31.9~$\pm$~3.9 &2.2491 & 73.3~$\pm$~2.2 \\
3.5761 & 26.0~$\pm$~3.4 &2.2503 & 65.7~$\pm$~1.9 \\
3.5776 & 16.3~$\pm$~1.3 &2.2514 & 45.7~$\pm$~1.4 \\
3.5790 &  7.7~$\pm$~1.4 &2.2526 & 22.2~$\pm$~0.7 \\
\end{tabular}
\end{ruledtabular}
\end{table}

\begin{figure}[t]
 \parbox{0.5\textwidth}{ \vspace{-0.5cm}
\includegraphics[width=0.43\textwidth]{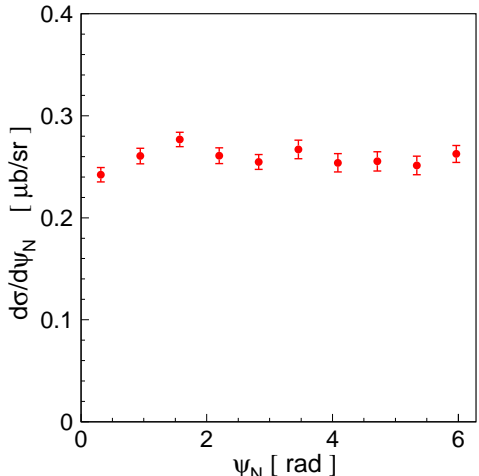}}
\vspace{-1.0cm}
\caption{\label{figurepsiN} 
 Differential cross section in $\psi_{N}$ for the $pp\to pp\eta$ reaction
 measured at Q~=~15.5~MeV. The variable $\psi_{N}$ is defined 
 in section~\ref{choiceofobservables}. 
}
\end{figure}

For completeness we  calculated also a
distribution of the angle $\psi_{N}$ defining the orientation of the 
$pp\eta$ system within the emission plane.  
The result is shown in figure~\ref{figurepsiN} and the values of $\frac{d\sigma}{d\psi_{N}}$
are listed in table~\ref{tablepsiN}.

Up to now we presented invariant mass spectra of the proton-proton
and proton-$\eta$ systems and angular distribution for two sets of 
non-trivial angles which describe the orientation of ejectiles within the 
emission plane and the alignment of the plane itself,
namely ($cos(\theta^{*}_{\eta})$, $\psi$) and ($cos(\theta^{*}_{N})$, $\psi_{N}$).

\begin{figure}[h]
    \parbox{0.5\textwidth}{\vspace{-0.5cm}
    \includegraphics[width=0.3\textwidth]{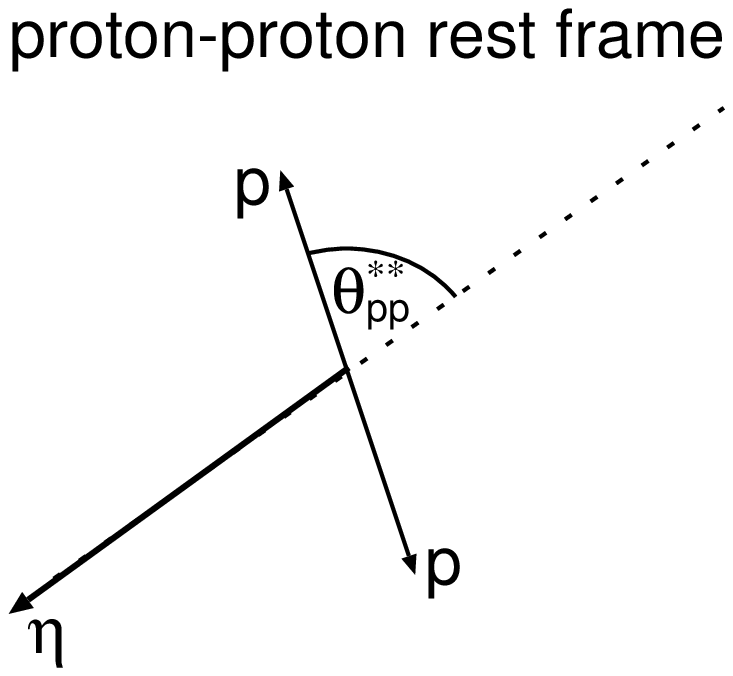}}
    \parbox{0.5\textwidth}{
    \includegraphics[width=0.4\textwidth]{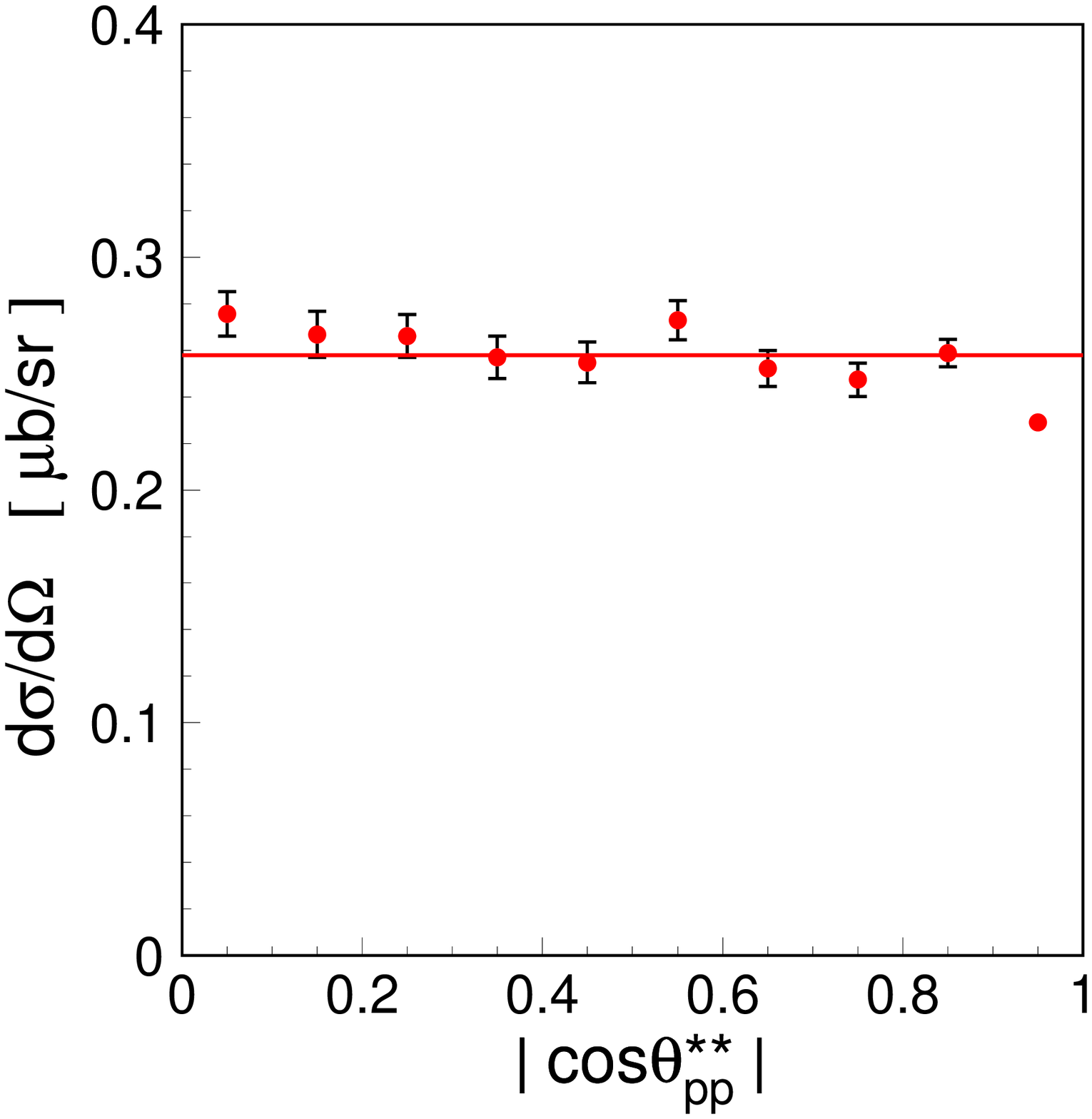}}\\
    \vspace{-0.6cm}
    \caption{\label{figureqpppp}
     Definition of $\theta^{**}_{pp}$, the polar angle of the relative
     proton-proton momentum with respect to the momentum of the $\eta$ meson
     as seen in the di-proton rest frame. 
     Lower picture shows differential cross section in $\theta^{**}_{pp}$
     as determined for the $pp\to pp\eta$ reaction at Q~=~15.5~MeV. 
    }
\end{figure}

Since one of the important issues which we will discuss in the next section is the 
contribution from  higher partial waves 
we evaluated also an angular distribution
of the relative momentum of two  protons seen from the proton-proton 
centre-of-mass subsystem~(see figure~\ref{figureqpppp}).
The distribution of that angle should deliver  information 
about the partial waves of the proton-proton system in the exit channel. 
In case of the two body scattering, the beam line,
which is at the same 
time the line along which the centre-of-mass system is moving, 
constitutes the reference frame for the angular distributions.
 For a three body final state
the beam axis is not a good direction 
to look for the angular distributions
\begin{table}[h]
\caption{\label{tablepsiN}
 Differential cross section in $\psi_{N}$ for the $pp\to pp\eta$ reaction
 measured at Q~=~15.5~MeV. The variable $\psi_{N}$ is defined 
 in section~\ref{choiceofobservables}.  Values are presented in figure~\ref{figurepsiN}.
}
\begin{ruledtabular}
\begin{tabular}{lr}
$\displaystyle{\psi_{N} [rad]}$ 
& $\displaystyle{\frac{d\sigma}{d\psi_{N}}}$ $\displaystyle{\left[\frac{\mu b}{sr}\right]}$ \\
\hline
0.314 &0.242~$\pm$~0.007 \\
0.942 &0.261~$\pm$~0.008 \\
1.571 &0.277~$\pm$~0.007 \\
2.199 &0.261~$\pm$~0.008 \\
2.827 &0.255~$\pm$~0.007 \\
3.456 &0.267~$\pm$~0.009 \\
4.084 &0.254~$\pm$~0.009 \\
4.712 &0.255~$\pm$~0.009 \\
5.341 &0.251~$\pm$~0.009 \\
5.969 &0.263~$\pm$~0.008 \\
\end{tabular}
\end{ruledtabular}
\end{table}
\begin{figure}[h]
\parbox{0.5\textwidth}{ \vspace{-0.3cm}
\includegraphics[width=0.40\textwidth]{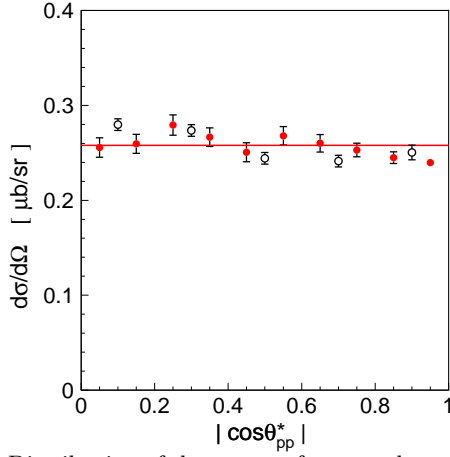}}
\vspace{-0.9cm}
\caption{\label{figureqpp} 
 Distribution of the centre-of-mass polar angle of the relative protons momentum
 with respect to the beam direction
 determined  for the $pp\to pp\eta$ reaction at Q~=~15.5~MeV.
 The COSY-11 result (closed circles) is compared to the data points determined 
 at Q~=~15~MeV by the TOF collaboration (open circles)~\cite{TOFeta}.
}
\end{figure}
relevant for the relative angular momenta of the two particles~\cite{kilian}. 
But by analogy to the two body system, an instructive reference axis 
for angular distributions in the proton-proton subsystem  
is now the momentum of the recoil $\eta$ meson, since the direction of that 
meson is identical with the direction of the movement of the proton-proton
centre-of-mass subsystem. 
The distribution of the differential cross section in $cos(\theta^{**}_{pp})$
is given in table~\ref{tablecosinusy}. In this table we listed also the 
differential cross section  as a function of angle $\theta_{pp}^{*}$ 
of the relative proton 
momentum seen from the overall centre-of-mass frame, 
as this is often considered in the theoretical works. 
In figure~\ref{figureqpp} our results are compared to the angular distribution
extracted by the TOF collaboration. Both experiments agree very well within the 
statistical accuracy, and both indicate a slight decrease of the cross section 
with increasing $|cos(\theta^{*}_{pp})|$.
\section{Interpretation of results}
\label{interpretation}
\begin{figure}
    \parbox{0.5\textwidth}{ \vspace{-1.3cm}
    \includegraphics[width=0.5\textwidth]{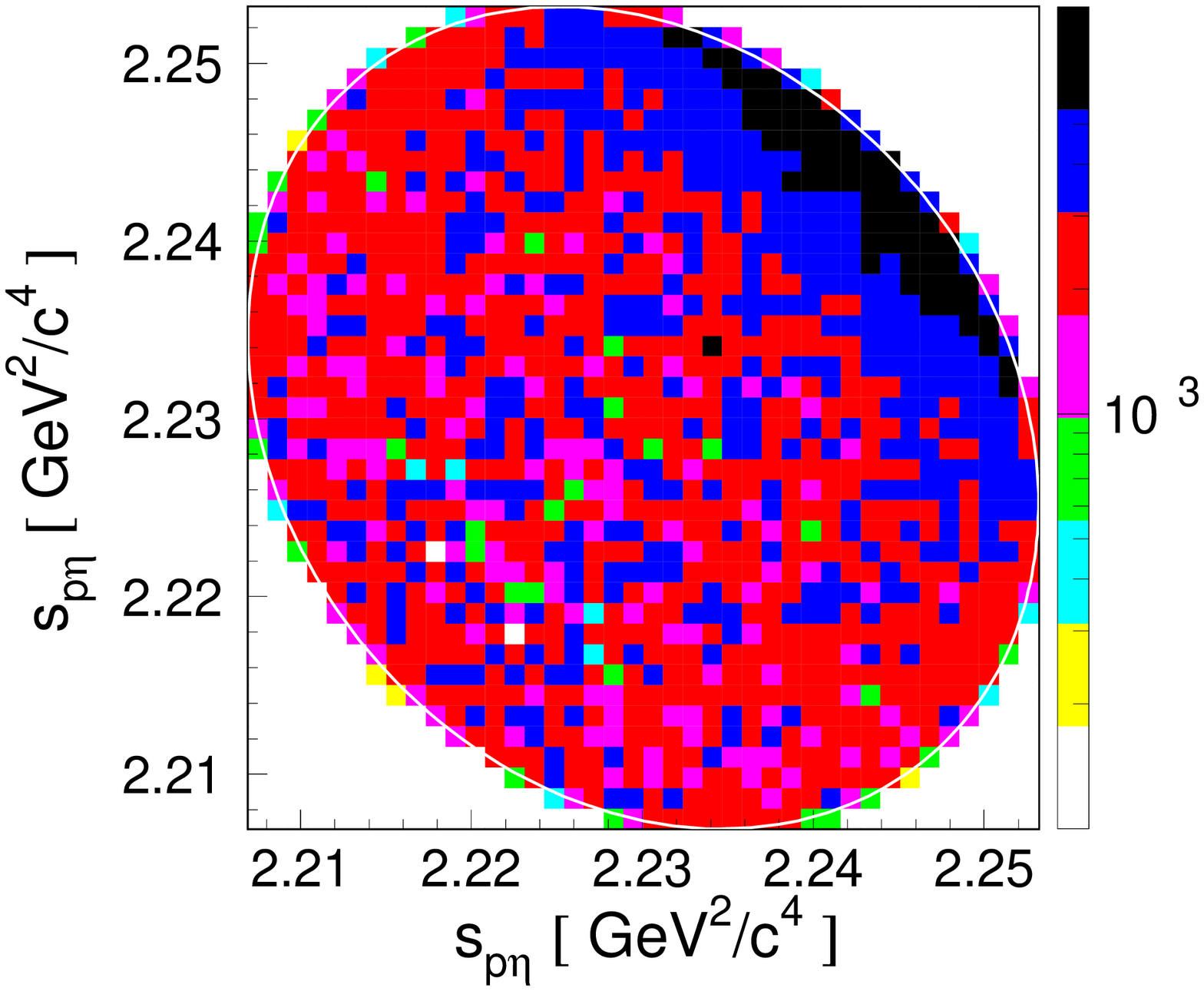}}
    \parbox{0.5\textwidth}{\vspace{-1.6cm}
    \includegraphics[width=0.5\textwidth]{dalitz_mc_gen_ppcfs_hab.eps}}
    \parbox{0.5\textwidth}{\vspace{-1.6cm}
    \includegraphics[width=0.5\textwidth]{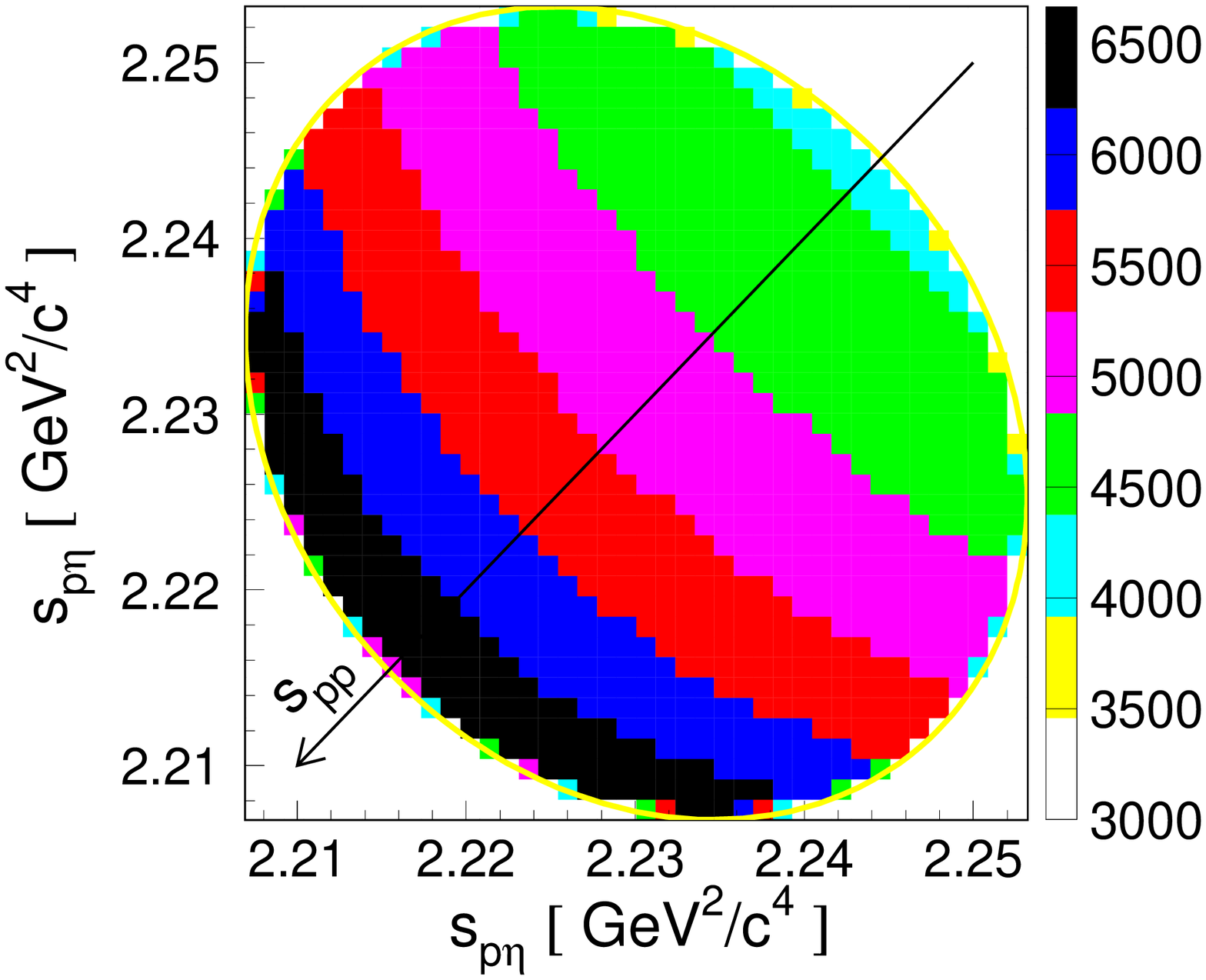}}
    \vspace{-0.8cm}
    \caption{\label{dalitze} Dalitz plot distributions. \\
     {\bf (upper panel)} Experimental result
       determined for the $pp\to pp\eta$ reaction at Q~=~15.5~MeV.
      Data were corrected for the detection acceptance and efficiency. \\
     {\bf (middle)}
      Monte-Carlo simulations for the $pp \to pp\eta$ reaction at Q~=~15.5~MeV:
      Phase-space density distribution modified by the 
      $^1S_{0}$ proton~-~proton final state interaction. \\
     {\bf (lower panel)} 
     Simulated phase-space density distribution modified by the proton-$\eta$ interaction
      with a scattering length of $a_{p\eta}$~=~0.7~fm~+~$i~0.3$~fm.
      Details of the calculations together with the discussion of the nucleon-nucleon
      and nucleon-meson final state interaction can be found in reference~\cite{review}.\\
      The scale in the figure is linear in contrast to the above panels.
      Lines surrounding the plots show the kinematical limits. 
    }
  \end{figure}

The interaction between particles depends on their relative momenta
or equivalently on the invariant masses of the two-particle subsystems.
Therefore it should show up as  modification
of the phase-space abundance in the kinematical regions where the outgoing
particles possess small relative velocities.
Only two  invariant masses of the three subsystems are independent
and therefore the entire  accessible information about the final
state interaction of the three-particle system can be presented in the
form of the Dalitz plot. The upper panel of figure~\ref{dalitze} 
indicates the event distribution
as determined experimentally for the $pp\eta$
system at an excess energy of Q~=~15.5~MeV. In this figure one easily recognizes
the growths of the population density at the region where the protons
have small relative momenta which can be assigned to the strong attractive
S-wave interaction between the two protons. This is qualitatively in agreement with the
expectation presented in figure~\ref{dalitze}(middle), which shows the results
of Monte-Carlo calculations  where the homogeneously populated
phase space was weighted by the square of the on-shell $^1S_{0}$
proton-proton scattering amplitude. However, already in this two dimensional
representation it is visible that
the experimentally determined distribution remains rather homogeneous
outside the region of the small proton-proton invariant masses,
whereas the simulated abundance decreases gradually with growing s$_{pp}$
(as indicated by the arrow).
The lower panel of figure~\ref{dalitze} shows the simulated  phase-space density
distribution
disregarding the proton--proton interaction but accounting for the interaction
between the $\eta$--meson and the proton.
Due to the lower strength of this interaction the expected deviations from a
uniform distribution is smaller by about two orders of magnitude, yet
an enhancement of the density in the range of low invariant
masses of proton--$\eta$ subsystems is clearly visible. Note that the scale
in the lower figure is linear whereas in the  middle and upper panel 
it is logarithmic.
Due to weak variations of the proton--$\eta$ scattering amplitude the
enhancement originating from the $\eta$--meson interaction with one proton is
not separated from the $\eta$--meson interaction with the second proton.
Therefore an overlapping of broad structures occurs.
It is observed that the occupation density grows slowly with increasing
$\mbox{s}_{pp}$, opposite to the effects caused by the S--wave proton--proton
interaction, but similar to the modifications expected for the NN P--wave~\cite{dyringPHD}.
From the above example it is obvious that, only
from experiments with high statistics,
signals of the meson--nucleon interaction can be observed  over the overwhelming
nucleon--nucleon final state interaction. 
  \begin{figure}
    \parbox{0.5\textwidth}{ 
    \includegraphics[width=0.5\textwidth]{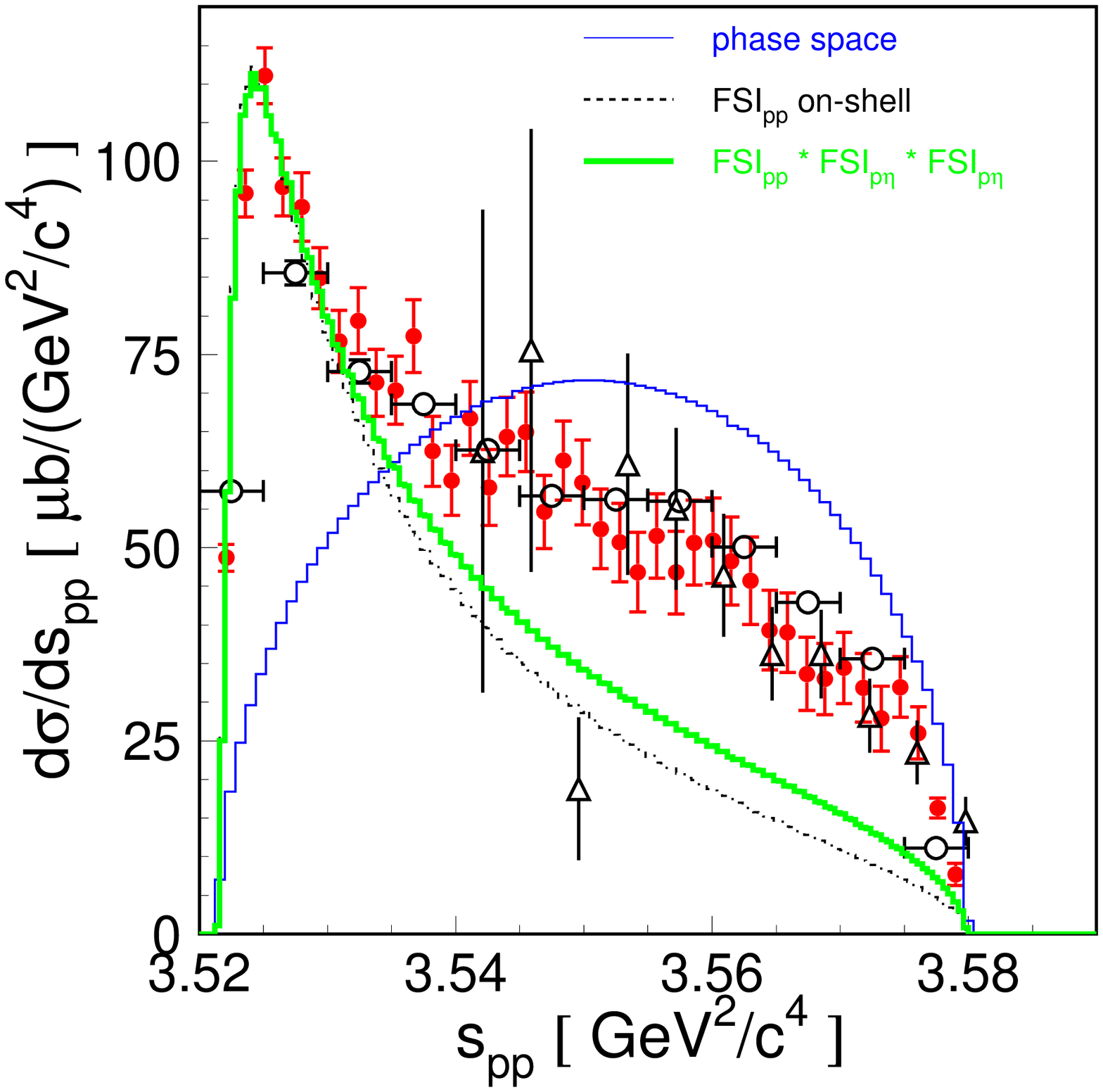}}
    \parbox{0.5\textwidth}{\vspace{-0.6cm}
    \includegraphics[width=0.5\textwidth]{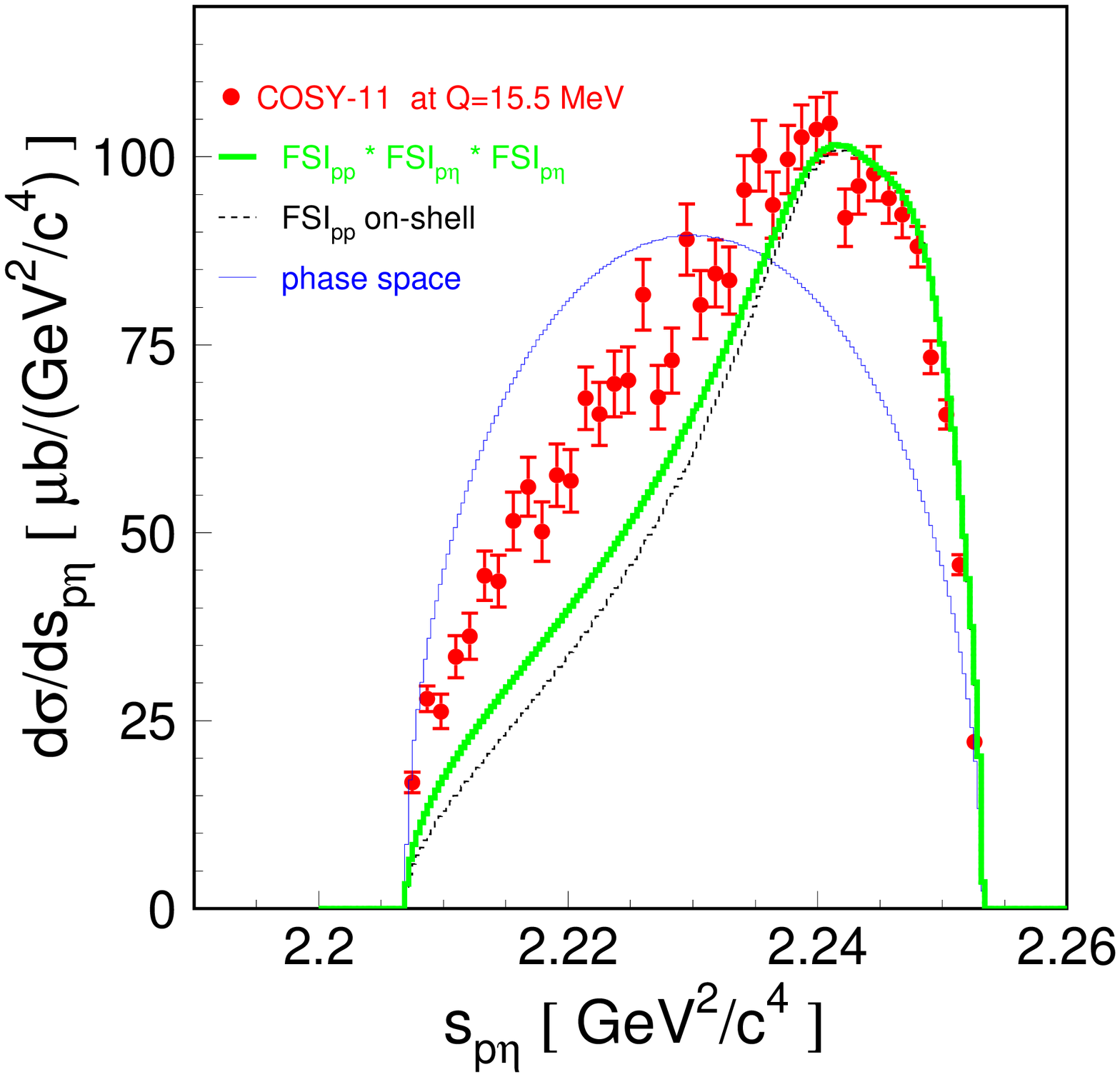}}
    \vspace{-0.6cm}
    \caption{\label{petaspp}
         Distributions of the square of the proton-proton~($s_{pp}$)
          and proton-$\eta$~($s_{p\eta}$)
          invariant masses
          determined experimentally
          for the $pp  \to pp\eta$ reaction at the excess energy of Q~=~15.5~MeV
          by the COSY-11 collaboration (closed circles),
          at Q~=~15~MeV by TOF collaboration (open circles)~\cite{TOFeta},
          and at Q~=~16~MeV by PROMICE/WASA (open triangles)~\cite{calen190}.
          The TOF and PROMICE/WASA data have been normalized
          to those of  COSY-11, since these  measurements 
          did not evaluate the luminosities but rather normalized 
          the results to  reference~\cite{caleneta} (see also comment~\cite{TOTAL}).
    The integrals of the phase space weighted by
    the square of the proton-proton on-shell
    scattering amplitude~(dotted lines)--FSI$_{pp}$, and by the product of FSI$_{pp}$ and
    the square of the proton-$\eta$ scattering amplitude~(thick solid lines),
    have been normalized arbitrarily at small values of $s_{pp}$.
    The thick solid line was obtained assuming a scattering length of
   $a_{p\eta}$~=~0.7~fm~+~$i$~0.4~fm.
    The expectation under the assumption of the homogeneously populated phase space
    are shown as thin solid curves.
    } 
  \end{figure}
 A deviation of the experimentally observed  population of the phase-space
 from the expectation based on the mentioned assumptions
 is even better visible in figure~\ref{petaspp}.
 This figure presents the  projection
 of the phase-space distribution onto the $s_{pp}$ axis corresponding
 to the axis indicated
 by the arrows in the two lower parts of figure~\ref{dalitze}.
 The superimposed lines in figure~\ref{petaspp} correspond to the calculations
 performed under the assumption that the production amplitude can be factorized
 into a primary production and final state interaction.
 The dotted lines result from calculations where only the proton-proton FSI was taken into account,
 whereas the  thick-solid lines present results where the overall enhancement
 was factorized into the corresponding pair interactions of the $pp\eta$ system.
 This factorisation Ansatz is of course only valid if the different amplitudes
 are completely decoupled which is certainly not the case here.
 Therefore, 
 this calculations should be considered as a rough estimate of the effect introduced
 by the FSI in the different two body systems.
 The enhancement factor accounting
 for the proton-proton FSI has been  calculated~\cite{swave,review}
 as the square of the on-shell proton-proton scattering amplitude
 derived according to the modified Cini-Fubini-Stanghellini  formula including
 the Wong-Noyes Coulomb corrections~\cite{noyes995}.
 The homogeneous phase-space distribution (thin solid lines)
 deviate strongly from
 the experimentally determined spectra.
 The curves including the  proton-proton and proton-$\eta$
 FSI reflect the
 shape of the data for small invariant masses of the proton-proton system,
 yet they  deviate significantly for large $s_{pp}$ and small $s_{p\eta}$ values.
  An explanation for this discrepancy 
  could be  a
  contribution from P-wave proton-proton 
  interaction~\cite{nakayamaa}, 
  or a
  possibly inadequate assumption
  that proton-$\eta$ and proton-proton interaction modify the
  phase space occupations only as
  incoherent weights~\cite{kleefeld}.
 
\begin{figure}
    \parbox{0.5\textwidth}{ 
    \includegraphics[width=0.5\textwidth]{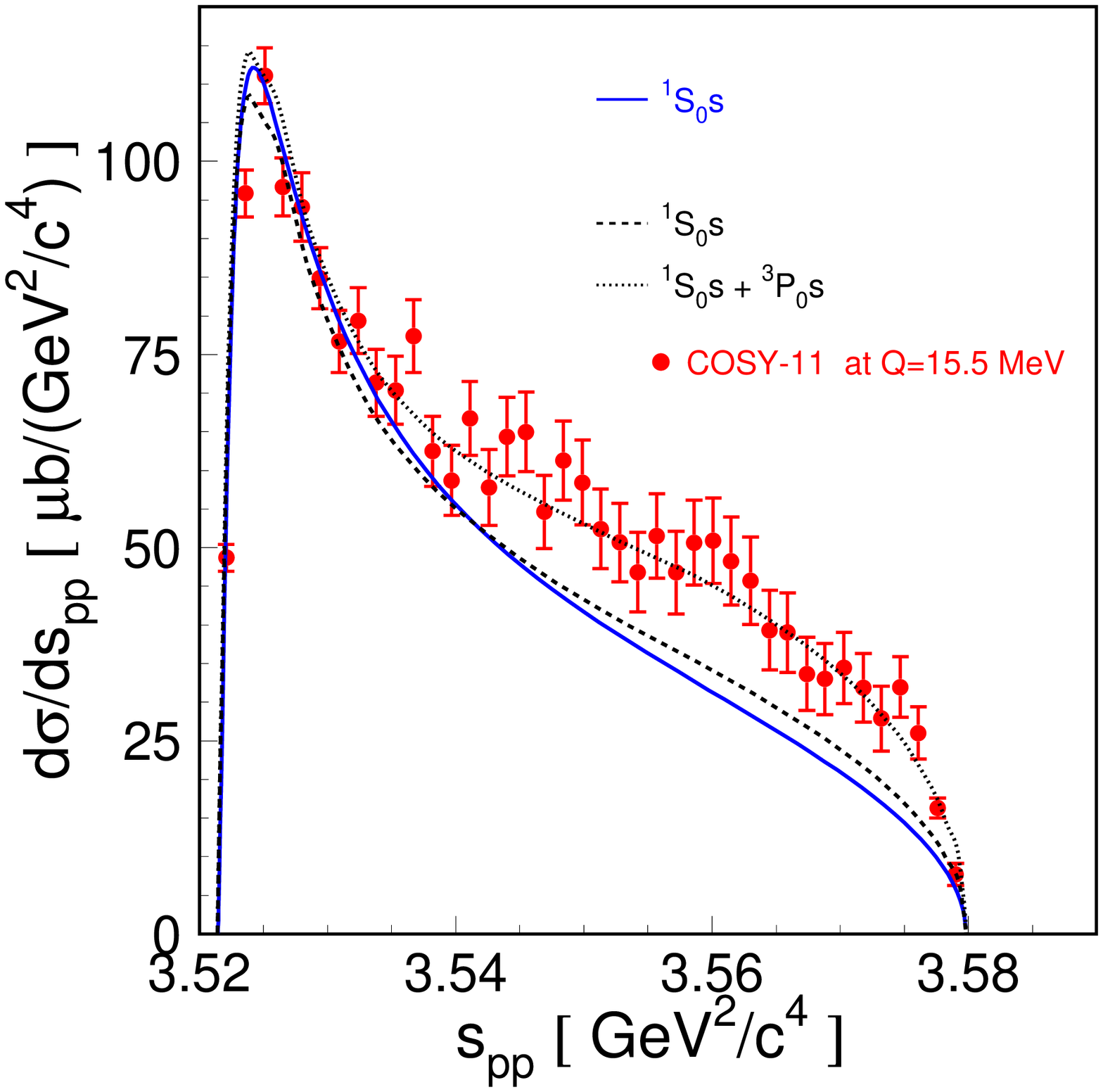}}
    \parbox{0.5\textwidth}{\vspace{-0.6cm}
    \includegraphics[width=0.5\textwidth]{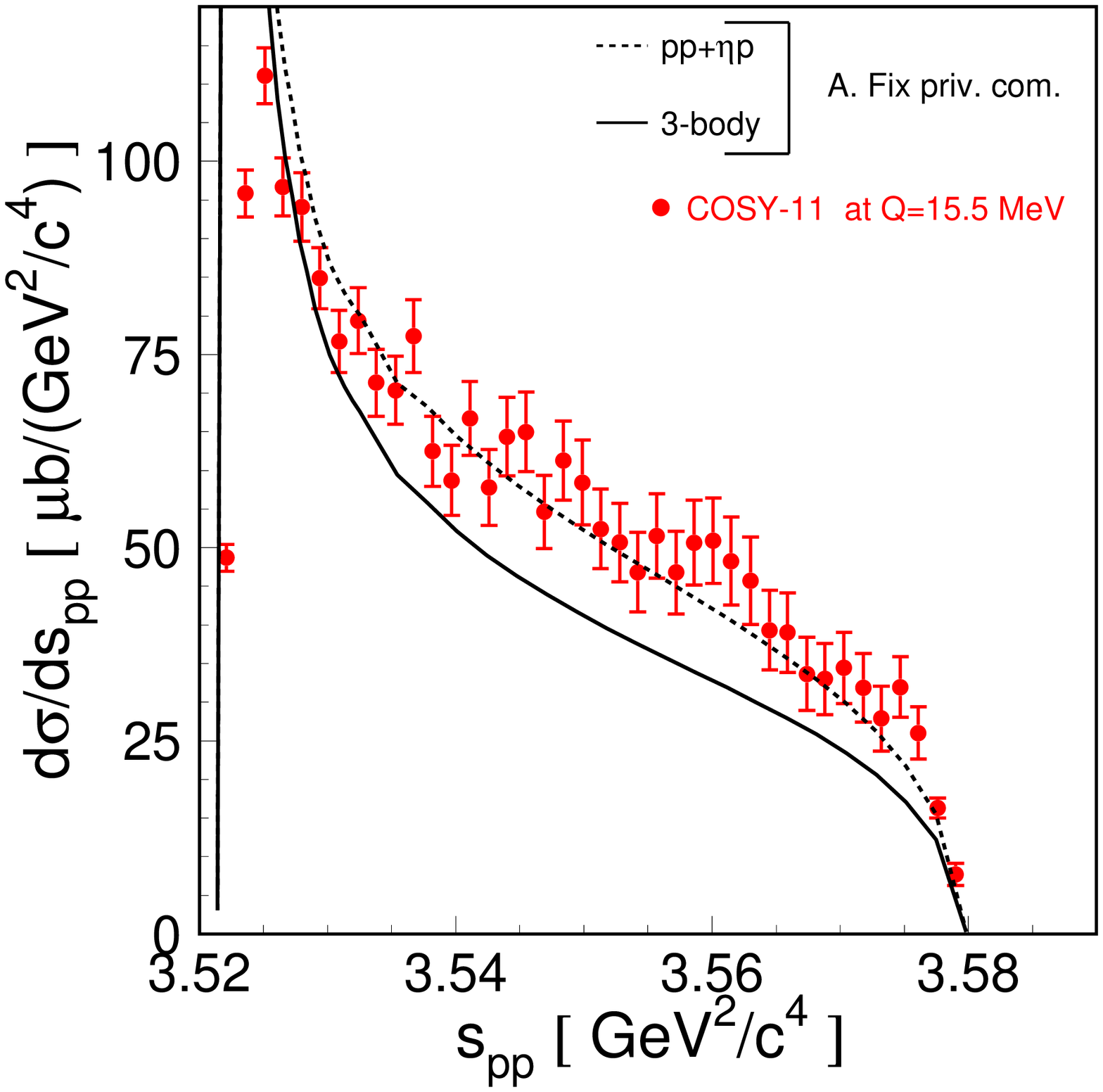}}
    \vspace{-0.6cm}
    \caption{\label{petaspp_kanzo}
       {\bf (upper picture)}
       Distribution of the square of the proton-proton~($s_{pp}$)
       invariant mass
       for the $pp\to pp\eta$ reaction at an excess energy of Q~=~15.5 MeV.
       Solid and  dashed lines corresponds to the calculations under the assumption
       of the $^3P_0\to ^1\!\!S_0s$ transition according to the models described in references~
       \cite{vadimm,vadimmpriv} and \cite{nakayamaa,nakayamaaa}, respectively.
       The dotted curve shows the result with the inclusion 
       of the  $^1S_0\to ^3\!\!P_0s$ contribution
       as suggested in reference~\cite{nakayamaa}. \\
       {\bf (lower picture)} The same data as above but with 
       curves denoting preliminary three-body calculations~\cite{Fixprivate} of the 
       final $pp\eta$ system as described in~\cite{fix3body}.
       At present only the dominant transition  $^3P_0\to ^1\!\!S_0s$
       is taken into account and
       the production mechanism is reduced to the excitation of the
       $S_{11}(1535)$ resonance via the exchange of the $\pi$ and $\eta$ mesons. 
       The solid line was determined with the rigorous three-body approach~\cite{Fixprivate}
       where the proton-proton sector is described in terms of the separable
       Paris potential (PEST3)~\cite{haidenbauer1822}, and for the $\eta$-nucleon scattering
       amplitude an isobar model analogous to the one of 
       reference~\cite{benhold625} is used
       with $a_{\eta N}~=~0.5$~fm~$+~i\cdot 0.32$~fm.
       The dashed line is obtained if only pairwise interactions ($pp+p\eta$) are allowed.
       The effect of proton-proton FSI at small $s_{pp}$ is overestimated due to neglect
       of Coulomb repulsion between the protons.
       The lines are normalized arbitrarily but their relative amplitude is fixed from the model.
    } 
  \end{figure}

   Slightly better description is achieved when
  the  proton-proton interaction is accounted for by the
  realistic nucleon-nucleon potential.
  The upper picture in figure~\ref{petaspp_kanzo} depicts the results obtained using
  two different models for the  production process as well as for the 
  NN interaction~\cite{vadimm,vadimmpriv}~\cite{nakayamaa,nakayamaaa}.
 The calculations for the $^3P_{0} \to ^1\!\!S_{0}s$ transition differ slightly,
  but the differences between the models are, by far, smaller than the
  observed signal.

  Therefore we can safely claim that the
  discussed effect is rather too large to be caused
  by the particular assumptions used for the production operator and
  NN potential. 
 
  As can be seen in figure~\ref{costhetaetatof},
  the distribution of the $\eta$ polar angle in the center of mass frame
  is fully isotropic.
  This is the next evidence ---~besides the shape of the excitation function
  and the kinematical arguments
  discussed in reference~\cite{review}~--- that
  at this excess energy (Q~=~15.5~MeV) the $\eta$ meson is produced
  in the center of mass frame predominantly with the angular momentum
  equal to zero. Similarly, the distribution determined for the
  polar angle of the relative
  proton-proton momentum with respect to the momentum of the $\eta$ meson
  as seen in the di-proton rest frame is also consistent with isotropy.
  Anyhow, even the isotropic distribution in this angle does not imply
  directly that the relative angular momentum between protons is equal to zero,
  because of their internal spin equal to $\frac{1}{2}$. 
  Therefore,
  the contribution from  the $^3$P$_0$-wave 
  produced via
  the $^1S_{0}\to ^3\!\!P_{0}s$ transition
  cannot be excluded. The isotropic angular distribution,
  as pointed out in reference~\cite{nakayamaa}, can
  also be principally achieved by the destructive interference between the
  transitions $^1S_{0} \to ^3\!\!P_{0}s$ and $^1D_{2} \to ^3\!\!P_{2}s$.
 
  As  show in reference~\cite{nakayamaa} the
  invariant mass distributions can be very well described when
  including higher partial-wave amplitudes.
  In fact, as depicted by the dotted line in the upper panel of figure~\ref{petaspp_kanzo},
  an admixture of the $^1S_{0}\to ^3\!\!P_{0}s$ transition leads to the
  excellent agreement with the experimentally determined
  invariant mass spectra.
  However, at the same time, the model of reference~\cite{nakayamaa} leads
  to  strong discrepancies in the shape of the excitation function
  as can be deduced from the comparison of the dashed-dotted
  line and the data in figure~\ref{cross_eta_etap}.
  Whereas it describes the data points
  in the excess energy range
  between 40~MeV  and 100~MeV it underestimates the total cross section
  below 20~MeV by a factor of 2.
\begin{table}[t]
\caption{\label{tablejurek}
  Square of the proton-proton invariant mass from the $pp\to pp\eta$ reaction
  measured for the excess energy range
  4~MeV~$\le$~Q~$\le$~5~MeV~\cite{jurekraport,smyrskieta}. 
}
\begin{ruledtabular}
\begin{tabular}{cc}
$\displaystyle{s_{pp} [GeV^2/c^4] }$ &  
$\displaystyle{\frac{d\sigma}{ds_{pp}} \left[\frac{\mu b}{GeV^2/c^4}\right]}$  \\
\hline
 3.52164 & 19.0~$\pm$~13   \\
 3.52204 & 74.0~$\pm$~21   \\
 3.52264 & 56.9~$\pm$~15   \\
 3.52344 & 91.7~$\pm$~17   \\
 3.52444 & 69.8~$\pm$~13   \\
 3.52564 & 56.9~$\pm$~11   \\
 3.52704 & 51.2~$\pm$~9.9  \\
 3.52864 & 41.8~$\pm$~8.4  \\
 3.53044 & 43.4~$\pm$~8.1  \\
 3.53244 & 36.6~$\pm$~7.0  \\
 3.53464 & 24.7~$\pm$~5.5  \\
 3.53704 & 6.8~$\pm$~2.8   \\
 3.53964 & 1.1~$\pm$~1.1   \\
\end{tabular}
\end{ruledtabular}
\end{table}

\begin{figure}[t]
\parbox{0.5\textwidth}{\vspace{-0.5cm}
\includegraphics[width=0.45\textwidth]{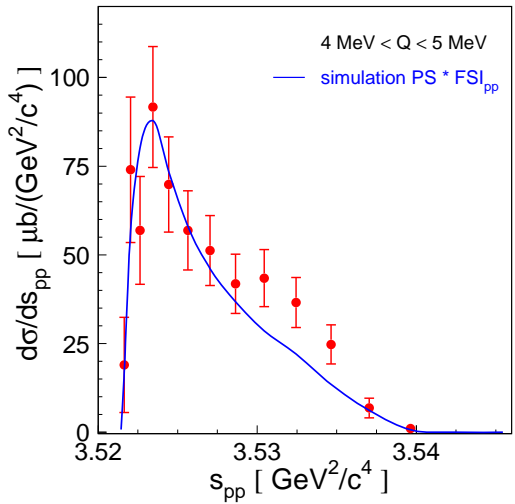}}
\vspace{-0.7cm}
\caption{\label{figurejurek} 
 Distribution of square of the proton-proton invariant mass from the $pp\to pp\eta$ reaction
  measured at COSY-11 for the excess energy range
   4~MeV~$\le$~Q~$\le$~5~MeV~\cite{jurekraport,smyrskieta}.
  Numerical values are listed in table~\ref{tablejurek}.
  The superimposed line shows the result of simulations performed under the assumption that 
  the phase space population is determined exclusively by the on-shell interaction 
  between outgoing protons. 
  The "tail" at large $s_{pp}$ values is due to the smearing of the excess energy,
  since this former COSY-11 data have not been kinematically fitted.
  Additionally to the 1~MeV range of Q a smearing of about 0.3~MeV~($\sigma$) should be taken 
  into account.
}
\end{figure}
  Interestingly, the enhancement at large $s_{pp}$ is visible
  also at much lower excess energy. This can be concluded from figure~\ref{figurejurek}
  in which the COSY-11 data at  Q~$\approx$~4.5~MeV~\cite{jurekraport}
  are compared to the simulations
  based on the assumption that the phase-space abundance is due to  
  the proton-proton FSI only.
   This observation could imply that the effect is caused by the 
  proton-$\eta$ interaction
  rather than  higher partial waves, since their contribution
  at such small energies is quite improbable~\cite{review}.
  However, as shown in the lower part of figure~\ref{petaspp_kanzo}
  the rigorous three-body treatment of the $pp\eta$ system leads
  at large values of $s_{pp}$ 
  to the reduction of the 
  cross section  in comparison to the calculation   
  taking into account only first order rescattering (pp+p$\eta$)~\cite{Fixprivate}.
  Here, both calculations include only the $^3P_{0}\to ^1\!\!S_{0}s$ transition.
  Though the presented curves are still preliminary, we can qualitatively assess  
  that the  rigorous three-body approach, in comparison to the present estimations,
  will on the one hand enhance the total cross section 
  near threshold as shown in figure~\ref{cross_eta_etap}, while on the other hand
  it will 
  decrease the differential cross section at large values of $s_{pp}$.
  This is just opposite to the influence  of P-waves 
  in the proton-proton system.  

  From the above presented considerations it is rather obvious 
  that the rigorous three-body treatment of the produced $pp\eta$ system
  and the exact  determination of the contributions from the 
  higher partial waves may result in the simultaneous explanation of  
  both observations:  the  near-threshold enhancement of the excitation function  
  of the total cross section, and the strong increase of 
  the invariant mass distribution at large values of $s_{pp}$.
  For the unambiguous determination of the contributions
  from different partial waves spin dependent observables are required~\cite{nakayamaa}.
  The first attempt has been already reported in~\cite{winter}.

\section{conclusion}
  Using the stochastically cooled proton beam at the Cooler Synchrotron COSY
  and the COSY-11 facility we have determined the total and differential 
  cross sections for the $pp\to pp\eta$ reaction at  an excess energy of Q~=~15.5~MeV.
  The high statistics data sample allowed us for the clear separation  of events
  corresponding to the $pp\to pp\eta$
  reaction from the multi-pion production 
  at each investigated phase-space bin, and  the multidimensional acceptance 
  correction allowed to extract the result without necessity of any assumption 
  about the reaction process. 

 The determined  distributions of the centre-of-mass polar angle of the
 $\eta$ meson emission as well as the distribution of the 
 relative proton-proton momentum with respect to the momentum of the $\eta$ meson
 are consistent with isotropy. Though, in the latter a small tendency
 of an increase of the cross section at 90 degree is observed.
 In contrary a rather strong decrease of the cross section was found 
 at 90 degree for the center-of-mass polar angle of the vector normal to the 
 emission plane. Explanation of that effect may reveal an interesting characteristic
 of the dynamics of the production process.
 
 The determined invariant mass spectra of the two-particle subsystems
 deviate strongly from the predictions based on the homogenous population
 of events over the phase space. Deviations at low proton-proton invariant
 mass values can be well explained as an influence of the S-wave interaction
 between the two protons. However, an unexpectedly large  enhancement of the
 occupation density in the kinematical regions of low $\eta$-proton relative
 momentum is not yet understood. 
 We have demonstrated that for the simultaneous description of the 
 excitation function and invariant mass distributions
 a rigorous three-body calculation with inclusion of the contribution 
 from higher partial waves is needed.

\begin{acknowledgments}
We acknowlegde the stimulating discussions and help of V.~Baru, 
A.~Fix and Ch.~Hanhart.
 The work has been partly supported by the European Community - Access to
Research Infrastructure action of the Improving Human Potential Programme
as well as the Internationales B\"uro and the Verbundforschung of the BMBF,
and by the Polish State Committe for Scientific Research (grant: 2P03B07123).
\end{acknowledgments}


\begin{thebibliography}{00}
\bibitem{nakayamaaa} 
 Here the model described in~\cite{nakayamaa} was 
 improved and the Coulomb interaction between protons was taken into account.
\end{thebibliography}
\end{document}